\documentclass{article}
\usepackage{arxiv}
\usepackage{color}
\usepackage{subfigure}
\usepackage{graphicx}
\usepackage{dcolumn}
\usepackage{bm}
\usepackage{braket}
\usepackage{floatrow}
\usepackage{amsmath}
\usepackage{makeidx}
\usepackage{epsfig}
\usepackage{array}
\usepackage{epstopdf}
\usepackage{multirow}
\usepackage[utf8]{inputenc} 
\usepackage[T1]{fontenc}    
\usepackage{hyperref}       
\usepackage{url}            
\usepackage{booktabs}       
\usepackage{amsfonts}       
\usepackage{nicefrac}       
\usepackage{microtype}      
\usepackage{lipsum}	
\usepackage{listofsymbols}
\usepackage{moreverb}
\usepackage{listings} 
\title{Theory and simulation of the Ising model}
\author{
  Ashkan Shekaari \\
  Department of Physics\\
  K. N. Toosi University of Technology\\
  Tehran, 15875-4416, Iran \\
  \texttt{shekaari@email.kntu.ac.ir} \\
   \And
 Mahmoud Jafari \thanks{Corresponding author} \\
  Department of Physics\\
  K. N. Toosi University of Technology\\
  Tehran, 15875-4416, Iran \\
  \texttt{jafari@kntu.ac.ir} \\}
\begin{document}
\maketitle
\begin{abstract}
We have provided a concise introduction to the Ising model as one of the most important models in statistical mechanics and in studying the phenomenon of phase transition. The required theoretical background and derivation of the Hamiltonian of the model have also been presented. We finally have discussed the computational method and details to numerically solve the two- and three-dimensional Ising problems using Monte Carlo simulations. The related computer codes in both Python and Fortran, as well as a simulation trick to visualize the spin lattice, have also been provided.
\end{abstract}
\section{Introduction}
Multifarious physical phenomena to which statistical mechanics~\cite{mk} is indeed applicable may fall into two major categories. In the first one, the system of interest could in fact be regarded as composed of practically noninteracting particles, in which the thermodynamic functions are straightforwardly calculated from the associated single-particle partition functions. Specific heats of gases and solids, chemical reactions and equilibrium constants, condensation of ideal Bose--Einstein gases, paramagnetism, spectral distribution of the blackbody radiation, and the elementary electron theory of metals, all belong to this category~\cite{1}. In solids, due to strong interatomic interactions, the actual atomic positions do not dramatically deviate from their mean values over a broad range of temperatures. As a result, a solid, in practice, could be thought of as an assembly of practically noninteracting harmonic oscillators (i.e., normal modes). In all these phenomena, except the Bose-Einstein condensation, the thermodynamic functions are smooth and continuous.

In the second category, on the other hand, the thermodynamic functions of the system in question involve analytic discontinuities and singularities in most cases, corresponding as well to various kinds of phase transitions. Condensation of gases, melting of solids, coexistence of phases particularly in the vicinity of critical points, ferromagnetism/antiferromagnetism, order-disorder transitions in alloys, and transitions to superconducting states, all fall into this category. The interparticle interactions featuring in these systems cannot be ignored, at all, by any approximation or transformation of the coordinates of the problem, in contrast to the first category aforementioned. As a result, the energy spectrum of the entire system cannot be related to its single-particle energy levels in any simple way. In such phenomena, a large number of corpuscles of the system may interact with one another in a rather strong, and cooperative manner, leading to phase transition as a macroscopic quality at a specific point called the critical temperature $(T_c)$ of the system.

It has been known that a simple-minded model consisting an array of lattice sites, with merely short-range, nearest-neighbor interactions, could mimic a number of physico-chemical systems in which phase transition takes place. Such a naive model is indeed good enough to understand a class of phenomena involving gas-liquid and liquid-solid transitions, ferromagnetism and antiferromagnetism, phase separation in binary solutions, and order-disorder transitions in alloys, on a unified, theoretical basis. Albeit dramatically oversimplified, this model retains the essential physical features of such systems, particularly those related to the propagation of long-range order in the systems. In the language of ferromagnetism, each of the $N$ lattice sites is considered to be occupied by an atom with a magnetic moment $\mathrm{\bm{\mu}}$ whose magnitude is $g\mu_{B}\sqrt{J(J+1)}$, which can accordingly take $(2J+1)$ allowed quantized orientations in space, giving rise to $(2J+1)^N$ different configurations for the entire system as well (bold characters denote vectors; $g$ is Lande's $g$-factor; $\mu_{B}\hspace{1.5mm}(=e\hbar\big/2mc)$ is the Bohr magneton; and $J$ is the eigenvalue of the total angular momentum operator). Each configuration (i.e., microstate) possesses an energy $E$ as a result of both pairwise interactions among neighboring atoms, and the interaction of the entire lattice with an external field $\mathrm{\bf{B}}$, if applied. The presence of a spontaneous magnetization $(M)$ at temperatures below $T_c$, and its absence above that temperature is then referred to as ferromagnetic phase transition. Theoretical and experimental investigations have so far shown that, for all ferromagnetic materials, data on the temperature dependence of $M$ fit best with the values of $J=1/2$ and $g=2.0023\cong 2$ in this model. As a result, it could be directly inferred that ferromagnetism is associated only with the spin degrees of freedom, not with the orbital motions of electrons~\cite{brnt,sct,sct2}. We then have $\mu=2\mu_{B}\sqrt{S(S+1)}$ in dealing with ferromagnetism, where $S$ is the quantum spin number (total spin). Moreover, $s=1/2$ (spin of an electron, or spin at a single lattice point) gives rise to only two orientations: $s_{z}=+1/2$ with $\mu_{z}=+\mu_{B}$, and $s_{z}=-1/2$ with $\mu_{z}=-\mu_{B}$, leading as well to $2^N$ configurations for the entire lattice.

The interaction energy between two neighboring spins $s_i$ and $s_j$, according to quantum mechanics, is given by $E_{ij}=K_{ij}\pm J_{ij}$, with the $(+)$ sign for antiparallel $(S=0)$ and the $(-)$ sign for parallel $(S=1)$ spins. The interaction energies $K_{ij}$ (Coulomb integrals)~\cite{par,ko} and $J_{ij}$ (exchange integrals)~\cite{ex} between particles $i$ and $j$ are also given by
\begin{equation*}
K_{ij}=\int\int\phi_{i}({\bm{\mathrm{r}}}_{1})\phi^{*}_{i}({\bm{\mathrm{r}}}_{1})\hat{u}_{ij}\phi_{j}({\bm{\mathrm{r}}}_{2})\phi^{*}_{j}({\bm{\mathrm{r}}}_{2})d{\bm{\mathrm{r}}}_{1}d{\bm{\mathrm{r}}}_{2},
\end{equation*}
and
\begin{equation*}
J_{ij}=\int\int\phi^{*}_{i}({\bm{\mathrm{r}}}_{1})\phi_{j}^{*}({\bm{\mathrm{r}}}_{2})\hat{u}_{ij}\phi_{i}({\bm{\mathrm{r}}}_{2})\phi_{j}({\bm{\mathrm{r}}}_{1})d{\bm{\mathrm{r}}}_{1}d{\bm{\mathrm{r}}}_{2},
\end{equation*}
where $\hat{u}_{ij}$ is the interaction potential (hat denotes operator), and $\phi_{i}({\bm{\mathrm{r}}}_{1})$ is the wavefunction of particle $i$ at position ${\bm{\mathrm{r}}}_{1}$---these integrals are both real and $K_{ij}\ge J_{ij}\ge 0$. The energy difference between the singlet ($E_s$) and triplet ($E_t$) states, for say a two-electron system, is also given by
\begin{equation}
\label{eq:e1}
E_{s}-E_{t}=2J,
\end{equation}
where $J\doteq J_{12}$ (see Appendix~\ref{sec:a1} for the derivation of Eq.~\ref{eq:e1}). The case with $J>0$ means the triplet state, which is energetically more favorable $(E_{t}=E_{s}-2J<E_{s})$, and the possibility of ferromagnetism should then be looked for. Antiferromagnetism, on the other hand, is marked by $J<0$, therefore $E_{s}=E_{t}-2J<E_{t}$. The Hamiltonian of the spin lattice can also be effectively written as
\begin{equation}
\label{eq:h}
\hat{H}=-2\sum_{i>j}J_{ij}\hat{s}_{i}.\hat{s}_{j},
\end{equation}
where $i$ and $j$ run over all spins, $\hat{s}_{i}$ is the spin of particle $i$, and the summation is carried out for all nearest-neighbor pairs in the lattice (see Appendix~\ref{sec:a2} for the derivation of Eq.~\ref{eq:h}). To a first approximation, we assume $J_{ij}$ to be considerable for only nearest-neighbor pairs (for which its value was denoted by $J$). This scheme, which is based on Eq.~\ref{eq:h}, is known as the Heisenberg model~\cite{wh}. Truncating $\hat{s}_{i}.\hat{s}_{j}=\hat{s}_{ix}\hat{s}_{jx}+\hat{s}_{iy}\hat{s}_{jy}+\hat{s}_{iz}\hat{s}_{jz}$ by only its last term (namely, $\hat{s}_{iz}\hat{s}_{jz}$), Eq.~\ref{eq:h} yields the so-called Ising model, first originated by Lenz~\cite{lnz} and followed up by his student Ising~\cite{isi}. As a result, Eq.~\ref{eq:h} turns into 
\begin{equation}
\label{eq:h2}
\hat{H}=-2J\sum_{i>j}\hat{\sigma}_{i}.\hat{\sigma}_{j},
\end{equation}
where the summation runs only over nearest-neighbor pairs, and $\hat{\sigma}$ is the Pauli matrix
$\begin{pmatrix} 
1 & 0 \\ 
0 & -1  
\end{pmatrix}$.
The truncation $\hat{s}_{i}.\hat{s}_{j}\cong\hat{s}_{ix}\hat{s}_{jx}+\hat{s}_{iy}\hat{s}_{jy}$ also gives rise to a different model used for quantum lattice gases, with possible relevance to superfluid transition in liquid He$^4$~\cite{mm}, as well as to the study of insulating ferromagnets. Both Ising and the so-called XY~\cite{bett} models could be regarded as special cases of a general anisotropic Heisenberg model with interaction parameters $J_{x,y,z}$: the former (Ising) corresponds with $J_{x},J_{y}\ll J_z$, while the latter represents $J_{x},J_{y}\gg J_z$. As mentioned before, the Ising model is capable of unifying phase transition phenomena occurring in several systems such as ferromagnets, gas-liquids, liquid mixtures, and binary alloys.

To further study and simulate the Ising model, we first ignore the kinetic energy of atoms and take into account only their spin degrees of freedom based on the fact that phase transition is essentially a consequence of the pairwise interaction. We then include only the nearest-neighbor contributions, as assumed in the model, in a way that the farther-neighbor contributions have no impact on the results. We could also study the relevant properties of the system under the effect of an external magnetic field $\mathrm{\bf{B}}$, causing in turn each spin $\sigma_i$ to acquire an extra potential energy $-\mu B\hat{\sigma}_i$ (where $B=|\mathrm{\bf{B}}|$). The full Hamiltonian of the system, using Eq.~\ref{eq:h2}, is then given by
\begin{equation}
\label{eq:h3}
\hat{H}=-2J\sum_{i>j}\hat{\sigma}_{i}.\hat{\sigma}_{j}-\mu B\sum_{i}\hat{\sigma}_{i}.
\end{equation}
\section{Theory}
\begin{figure}[H]
	\centering
	\fbox{\rule[0cm]{0cm}{0cm}     \rule[0cm]{0cm}{0cm}
	\includegraphics[scale=0.6]{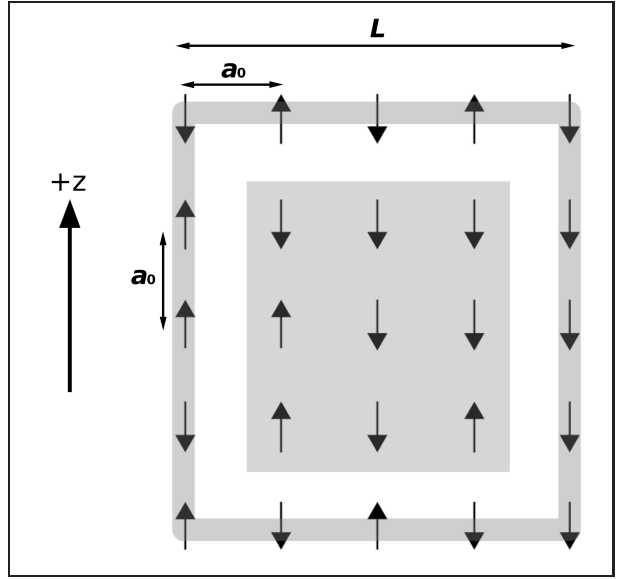}}
	\caption{\label{fig:1}
		A 2D, $5\times 5$, square lattice of identical spins with the lattice constant $a_0$ and the length $L$---rendered in GIMP (version 2.8)~\cite{gmp}. Taking $a_0$ to be the length scale, $1+L\big/a_0$ gives the number of spins in each row or column. Only two possible space orientations (namely, $\pm z$) are assumed to be allowed, meaning spin--\small{1/2} particles. The total number of spins is also $N=25$, out of which $N_{b}=16$ spins are on the border (indicated by the gray narrow band), and the rest, $N_{v}=9$ spins, are in the volume (the central gray area) of the system, showing that the ratio $N_{b}\big/N_{v}\simeq 1.77\gg 0$ is quite large, affecting dramatically the physics of the problem as well.}
\end{figure}

The Ising model has exact solutions in one and two dimensions, in contrast to the 3D cases all we know about which is through numerical simulations. Here, we aim at numerically solving the two- (2D) and three-dimensional (3D) Ising problems obeying Eq.~\ref{eq:h3} excluding the last term, namely with $B=0$ without loss of generality, using the Monte Carlo (MC) method~\cite{mc}. To this end, we assume that the spins have primarily been aligned along the $z$ direction, as illustrated in Fig.~\ref{fig:1} for a 2D, $5\times 5$, square lattice.
The minus sign behind $J$ in Eq.~\ref{eq:h3} indicates that any two neighboring spins prefer to be in the same direction. In other words, to maintain this minus, which is fundamental in order for the entire system to have a negative energy, both $\sigma_{i}$ and $\sigma_{j}$ must be either $+1$ or $-1$; otherwise, $\sigma_{i}.\sigma_{j}=-1$ leading as well to a positive total energy value for our bounded system. Therefore, any two neighboring spins tend to be parallel in either direction. It is also easy to infer that if the system is left to its own devices, namely in the absence of any external agitating factor such as magnetic fields (then $B=0$) or thermal fluctuations (then $T=0$), all spins tend to be parallel in either space orientation, and the minimum-energy $(E_{min})$ configuration is accordingly obtained. In such a case, the minimum number of microstates of the system is evidently $\Omega_{min}=2$ (all spins up or down), leading in turn to minimum entropy $(\mathcal{S}_{min}=k_{B}\ln\Omega_{min}=k_{B}\ln 2$; $k_{B}$ being the Boltzmann constant). Moreover, because $T=0$, minimum free energy $(F_{min})$ is also obtained according to the thermodynamic relation $F=E-T\mathcal{S}$. Evidently, any difference in the sign of neighboring spins increases the energy of the system. 

Nevertheless, for $T>0$, minimum energy is not concurrent with minimum free energy due to $T\mathcal{S}>0$. To further clarify this point, we consider two microstates of the lattice of Fig.~\ref{fig:1}: one with all spins up, and the other with only one spin down. For the former, $\Omega=1$ and then $\mathcal{S}=k_{B}\ln 1=0$. For the latter, however, $\mathcal{S}=k_{B}\ln\left[N!\big/(N-1)!\right]=k_{B}\ln 25$, according to the fact that the only opposite (down) spin can be in any lattice point. The microstate with half of the spins up (or down) thus possesses both maximum entropy and zero magnetization $(M)$ simultaneously. As a result, $F_{min}$ is obtained when $T\mathcal{S}$ is maximum, namely when $\mathcal{S}$ is maximum, or for high enough temperature values that again lead to maximum entropy, or to $M=0$. It means that at high enough temperatures, the system lacks a net magnetization, and vice versa, as temperature decreases, $T\mathcal{S}$ becomes small, $F$ approaches $E$, and the system accordingly possesses a net, non-zero magnetization. According to this inverse correlation between $M$ and $T\mathcal{S}$, the former can then be viewed as an indication of order of the system, called as well the order parameter. At a critical temperature ($T_c$), the system undergoes a phase transition in a way that $M$ suddenly drops to zero; for $T<T_c$ the system is in the state in which all the spins are either up or down (minimum entropy), while for $T>T_c$ the system asymptotically moves to the state with half of the spins up or down (maximum entropy). As a result, this model features one of the most important phenomenon in physics, which is called {\em{spontaneous symmetry breaking}}~\cite{ssb}: at temperatures considerably higher than $T_c$, there exists an up-down spin symmetry, which spontaneously breaks into either spins-up or spins-down state as temperature approaches $T_c$. 

Studying the behavior of the system in the vicinity of its critical point is indeed a basic problem in the theory of phase transition. Such a behavior is marked by the fact that various physical quantities of the system (e.g., the thermodynamic observables and\big/or their derivatives) possess singularities or discontinuities at the critical temperature. Notable examples of this type in the Ising model are the critical behaviors of the specific heat $(C_v)$ and the magnetic susceptibility $(\chi)$. From statistical mechanics, such singularities are direct consequences of the fact that the correlation length $(\xi)$ of the system becomes infinitely large at $T_c$. This quantity is in fact a manifestation of the correlation between spin positions in the lattice. Due to such an infiniteness, the system becomes scale-free meaning that the asymptotic behaviors of the associated physical observables of the system near the critical point are given by power laws---say $M\sim (T_{c}-T)^{\delta}$, $\chi\sim |T_{c}-T|^{-\gamma}$, $C_{v}\sim |T_{c}-T|^{-\alpha}$, and $\xi\sim |T_{c}-T|^{-\nu}$, where $|T_{c}-T|$ means both $T_{c}-T$ and $T-T_c$ depending on the sign of $T_{c}-T$ at the neighborhood of $T_c$. As is seen, these power laws are characterized by the parameters $\alpha,\gamma,\delta$, and $\nu$, known as critical exponents, determining as well the qualitative nature of the critical behavior of the system.
\subsection{Ensemble and phase space of the system}
Assume that the original system of interest is a 2D square lattice with a total number $N$ of localized identical spins. Based on the fact that the standard statistical ensemble in MC simulations is the canonical ensemble [this is indeed due to the natural appearance of the equilibrium temperature in the Boltzmann weight factor $e^{-\beta E}$ ($\beta=1/k_{B}T$)], we accordingly consider a large ensemble consisting of many mental copies of the original system, being naturally enough in all sorts of possible configurations (microstates), including as well all possible states.

Considering this ensemble, the various systems will be in all sorts of possible microstates at any instant of time, consistent with the given macrostate which is common to all of them. In the phase space of the system, the corresponding picture involves a swarm of representative phase points, each of which for one member of the ensemble. In terms of dynamics, as the systems of the ensemble move continually between different microstates with time, the corresponding representative points also simultaneously move along their respective phase-space paths.

In the most general case, supposing the original system of interest to be a $D$-dimensional lattice with $N$ identical spins localized on the lattice points, the average of the thermodynamic observable $A$ pertaining to this system in say the canonical ensemble is then given by
\begin{equation}
\label{eq:fr}
\langle A\rangle=\bigg[\int A(q,p)e^{-\beta E(q,p)}d^{DN}q\hspace{0.5mm}d^{DN}p\bigg]\Bigg/\bigg[\int e^{-\beta E(q,p)}d^{DN}q\hspace{0.5mm}d^{DN}p\bigg],
\end{equation}
which involves two many-dimensional integrals, $\big(d^{DN}q\hspace{0.5mm}d^{DN}p\big)$ is the volume element in the ($DN+DN=$) $2DN$-dimensional phase space, $q$ and $p$ are respectively the position and momentum coordinates, and $E(q,p)$ is the total energy of the system. Even, by taking into account the fact that interactions between the spins are velocity-independent (according to Eq.~\ref{eq:h3}) and the momentum integrals could accordingly be separated off, carrying out integration in a $DN$-dimensional configuration space using usual numerical techniques is still quite impossible---$e^{-\beta E(q,p)}$ is in general the density of the phase-space representative points, which can also be more complicatedly a function of time. Resorting to MC methods, Eq.~\ref{eq:fr} accordingly turns into the simple unweighted average
\begin{equation}
\label{eq:avg}
\langle A\rangle=\frac{1}{M}\sum_{i=1}^{M}A({\mathfrak{q}}_i)
\end{equation} 
where $M$ is the number of microstates ${\mathfrak{q}}_i$ within the sample $\{\mathfrak{q}\}$---a sample, here, is referred to as a set of microstates, which is in turn a subset of the ensemble.
\subsection{The modified Monte Carlo scheme}
In statistical mechanics, computer simulations involve two major categories, including molecular dynamics (MD)~\cite{md}, and MC. In equilibrium statistical mechanics, an MC simulation uses pseudorandom number generators to sample the ensemble of the system according to the equilibrium probability distribution
\begin{equation}
\label{eq:pr}
\mathcal{P}(\mathfrak{q})=\frac{e^{-\beta E(\mathfrak{q})}}{\sum_{\mathfrak{q'}}e^{-\beta E(\mathfrak{q'})}},
\end{equation}
where the summation is carried out over all members of the set $\{\mathfrak{q'}\}$. In a modified MC scheme devised by Metropolis {\em{et al.}}~\cite{mp}, an MC simulation samples configurations (microstates) according to the Boltzmann weight factor $e^{-\beta E}$ and then weights them evenly, instead of drawing a sample randomly and then weighting them by $e^{-\beta E}$. The former, referred to as {\em{importance sampling}}, is indeed more optimized and computationally cost-effective compared to the latter simple-minded random sampling, and brings about a remarkable computational efficiency because according to which, all microstates within the sample are chosen with a probability $e^{-\beta E}$. Therefore, expectation values of the thermodynamic observables could simply be calculated by Eq.~\ref{eq:avg}.
\subsection{The Metropolis algorithm}
Random walk in the sampling space, i.e., selecting a random microstate $\mathfrak{q}$ from the set $\{\mathfrak{q}\}$ with a probability distribution approaching Eq.~\ref{eq:pr}, is accomplished by the Metropolis algorithm. Using Eq.~\ref{eq:pr}, for two consecutive states $\mathfrak{q}$ and $\mathfrak{q'}$ (the system moves from the former to the latter state in its phase space), it is obtained that
\begin{equation*}
\frac{\mathcal{P}(\mathfrak{q'})}{\mathcal{P}(\mathfrak{q})}=e^{-\beta\big[E(\mathfrak{q'})-E(\mathfrak{q})\big]}=e^{-\beta\Delta E}.
\end{equation*}
This factor, as we show later on, lies at the heart of the Metropolis algorithm, and governs as well the random walks on the lattice. One important feature of this ratio is that it depends only on the energy difference between any two states, therefore, there is no need to have energies of all the microstates as included in the denominator of Eq.~\ref{eq:pr} as the canonical partition function of the system.

The Metropolis algorithm is accomplished in the following five steps: 
\begin{itemize}
\item Step I. Change the present microstate $\mathfrak{q}$ of the system to another randomly chosen state $\mathfrak{q'}$. In an Ising problem, this is equivalent to flipping a randomly selected spin. Since the total energy $(E)$ explicitly appears in the Boltzmann weight factor, it must then be an explicit function of the microstate of the system in a way that change in $\mathfrak{q}$ directly changes $E$. 

\item Step II. Calculate the energy difference $\Delta E=E(\mathfrak{q'})-E(\mathfrak{q})$. If $\Delta E\le 0$, then $E(\mathfrak{q'})\le E(\mathfrak{q})$ [therefore $\mathcal{P}(\mathfrak{q'})\ge\mathcal{P}(\mathfrak{q})$] which is a desired result due to bringing about a less total energy compared to the previous state of the system, the change must then be accepted. In contrast, $\Delta E>0$ is not desired, nevertheless, we do not ignore this case and accept the related change with probability $e^{-\beta\Delta E}$. And how one can accept a change with a probability? The most familiar way is undoubtedly to use a die. As an illustration, assume you aim to accept a change or make a selection with probability $P_0=0.25$. You accordingly have to design a tetrahedron, pyramid-shape die with four equilateral triangular faces. Indeed, when we use coins or cubic dice, we are making selections with probabilities 0.5 and 0.167, respectively. Here, because $\Delta E>0$, $e^{-\beta\Delta E}$ is always between 0 and 1, and we can then design a computational die with $e^{+\beta\Delta E}$ sides. Since this exponential almost never results in a uniform die due to being inherently any real number---in other words, choosing a (real) random number between 0 and 1 almost never would be equal to a specific value of $e^{-\beta\Delta E}$ due to the never-happening match between their decimals---a more general, less stringent condition is then to accept the change with a probability less than $e^{-\beta\Delta E}$. This condition can indeed be justified according to the fact that the change with $\Delta E>0$ is not a desired one and we basically tend to get rid of this. As a result, being $rand()$ a random number between 0 and 1 as a probability, if $rand()<e^{-\beta\Delta E}$, then the change is accepted. Albeit seemingly evident, it should be useful to note that producing one random number using a pseudorandom number generator is equivalent to one throw of our computational die.

\item Step III. Perform steps I and II once for each spin in the system (i.e., one spin at a time). The spins are usually selected randomly to ensure {\em{detailed balance}}~\cite{db,db2}. Steps I through III also define one MC sweep.

\item Step IV. Repeat steps I through III for a number (\texttt{eqstp} in the codes) of MC sweeps in order for the entire system to be equilibrated. This value cannot be indeed estimated {\em{a priori}}. As a result, for an arbitrary value of temperature in the temperature loop of the code, one has to plot the total energy of the system at each step of equilibration and then check whether convergence is achieved.

\item Step V. Repeat steps I through III for a number (\texttt{mcstp} in the codes of Sec.~\ref{sec:34}) of MC sweeps.
\end{itemize}
\subsection{Ergodicity}
In both major classes of computer simulation in statistical mechanics, MD and MC, the condition of being ergodic is cardinal, without which the simulations are not reliable to calculate the thermodynamic averages of interest. In the present MC algorithm, since all spins are equally likely to be chosen with finite probabilities, a large enough number of random moves would accordingly give the chance to all of them to be flipped, showing the ergodicity of the method. In other words, any microstate of the system can be reached from any other. More importantly, if we make a change in all the systems of the ensemble, and define $P_{\mathfrak{q}\rightarrow \mathfrak{q'}}$ as the probability of a system to be carried out from state $\mathfrak{q}$ to $\mathfrak{q'}$ caused by that change, ergodicity guarantees that 
\begin{equation}
\label{eq:dbal}
P_{\mathfrak{q}\rightarrow \mathfrak{q'}}=P_{\mathfrak{q'}\rightarrow \mathfrak{q}},
\end{equation}
because all possible microstates are equally likely to happen.
\subsection{Canonical distribution of the ensemble}
The initial distribution of the ensemble, involving many copies of the original system, is evidently random: ${\mathcal{M}}_{\mathfrak{q}}$ systems in state $\mathfrak{q}$, ${\mathcal{M}}_{\mathfrak{q'}}$ systems in state $\mathfrak{q'}$, ${\mathcal{M}}_{\mathfrak{q''}}$ systems in state ${\mathfrak{q''}}$, and so on. We indeed aim at approaching an ensemble in which the distribution of any configuration (microstate) ${\mathfrak{q}}_i$ is given by Eq.\ref{eq:pr}. The why of this goal is that the final ensemble will be accordingly composed of low-energy configurations, namely the states that are most probable to occur in Nature, including as well the ground state of the system. This, in fact, is a kind of importance sampling, and once the probability distribution of our ensemble approaches Eq.\ref{eq:pr}, we are then allowed to weight them evenly, which exactly means that the simulation averages of the thermodynamic observables are simply given by Eq.~\ref{eq:avg}.

We now assume that ${\mathcal{M}}_{\mathfrak{q}}$ is the number of systems of the ensemble that are in state $\mathfrak{q}$, and $E(\mathfrak{q})>E(\mathfrak{q'})$, $\mathfrak{q'}$ being another state. Therefore, transition from $\mathfrak{q}$ to $\mathfrak{q'}$ is allowed due to bringing about a lower energy, and the total number of systems moving from $\mathfrak{q}$ to $\mathfrak{q'}$ is simply given by ${\mathcal{M}}_{\mathfrak{q}}P_{\mathfrak{q}\rightarrow \mathfrak{q'}}$. The number of systems moving from $\mathfrak{q'}$ to $\mathfrak{q}$ is also
\begin{equation*}
\mathcal{N}(\mathfrak{q'}\rightarrow \mathfrak{q})={\mathcal{M}}_{\mathfrak{q'}}P_{\mathfrak{q'}\rightarrow \mathfrak{q}}e^{-\beta\Delta E}={\mathcal{M}}_{\mathfrak{q'}}P_{\mathfrak{q}\rightarrow \mathfrak{q'}}e^{-\beta[E(\mathfrak{q})-E(\mathfrak{q'})]},
\end{equation*}
obtained according to the Metropolis algorithm (step II), and by using Eq.~\ref{eq:dbal}. As a result, the net difference between the number of systems moving from $\mathfrak{q}$ to $\mathfrak{q'}$ and that of the reverse direction is then
\begin{equation*}
\mathcal{N}(\mathfrak{q}\rightarrow \mathfrak{q'})-\mathcal{N}(\mathfrak{q'}\rightarrow \mathfrak{q})={\mathcal{M}}_{\mathfrak{q}}P_{\mathfrak{q}\rightarrow \mathfrak{q'}}-{\mathcal{M}}_{\mathfrak{q'}}P_{\mathfrak{q}\rightarrow \mathfrak{q'}}e^{-\beta[E(\mathfrak{q})-E(\mathfrak{q'})]}=P_{\mathfrak{q}\rightarrow \mathfrak{q'}}\Big({\mathcal{M}}_{\mathfrak{q}}-{\mathcal{M}}_{\mathfrak{q'}}e^{-\beta[E(\mathfrak{q})-E(\mathfrak{q'})]}\Big).
\end{equation*}
From the Metropolis algorithm, it can be inferred that more systems will eventually move from state $\mathfrak{q}$ to $\mathfrak{q'}$ on average, according to our previous assumption $E(\mathfrak{q'})<E(\mathfrak{q})$, therefore,
\begin{eqnarray*}
\mathcal{N}(\mathfrak{q}\rightarrow \mathfrak{q'})>\mathcal{N}(\mathfrak{q'}\rightarrow \mathfrak{q})\Longrightarrow
\mathcal{N}(\mathfrak{q}\rightarrow \mathfrak{q'})-\mathcal{N}(\mathfrak{q'}\rightarrow \mathfrak{q})>0\Longrightarrow
{\mathcal{M}}_{\mathfrak{q}}-{\mathcal{M}}_{\mathfrak{q'}}e^{-\beta[E(\mathfrak{q})-E(\mathfrak{q'})]}>0\nonumber\\
\Longrightarrow
{\mathcal{M}}_{\mathfrak{q}}>{\mathcal{M}}_{\mathfrak{q'}}\frac{e^{-\beta E(\mathfrak{q})}}{e^{-\beta E(\mathfrak{q'})}}=\bigg(\frac{e^{-\beta E(\mathfrak{q'})}}{{\mathcal{M}}_{\mathfrak{q'}}}\bigg)^{-1}e^{-\beta E(\mathfrak{q})}
\Longrightarrow
{\mathcal{M}}_{\mathfrak{q}}=c_{0}\bigg(\frac{e^{-\beta E(\mathfrak{q'})}}{{\mathcal{M}}_{\mathfrak{q'}}}\bigg)^{-1}e^{-\beta E(\mathfrak{q})}
\Longrightarrow
{\mathcal{M}}_{\mathfrak{q}}\propto e^{-\beta E(\mathfrak{q})},
\end{eqnarray*}
meaning that the ensemble will approach the canonical distribution---here, $c_0$ is some constant, and the ratio with the exponent $(-1)$ is also a detailed average, namely the probability of being in microstate $\mathfrak{q'}$ averaged over all the systems that are in this state.
\section{Simulation details}
\subsection{Initialization}
The initial state of the system can be a random distribution of $(\pm 1)$ (associated with the two eigenvalues of $\hat{\sigma}_z$). It is also quite reasonable to start with a minimum-entropy configuration, such as the spins-up microstate, as we do here. We therefore store the number $N$ of $(+1)$s as elements of an array, which are then loaded into the computer memory once the code is run (one can also store them into an input file, but it is indeed redundant due to the fact that the initial configuration of the system after few MC sweeps is entirely lost and will not be reversible/retrievable anymore).
\subsection{Energy}
The total energy of the system, according to Eq.~\ref{eq:h3} (with $B=0$), is given by
\begin{equation}
\label{eq:et}
-E\Big/J=2\sum_{i>j}\sigma_{i}.\sigma_{j}=\sum_{i=1}\sum_{j=1}\sigma_{i}.\sigma_{j},
\end{equation}
according to which $J$ can be simply chosen as the unit of energy ($E\big/J$ is dimensionless). The energy difference is also calculated as follows. Assume in an MC sweep, the spin $s_{ij}$ is supposed to be flipped. Before flip (for a 2D lattice) we have
\begin{equation*}
E(\mathfrak{q})=E_{0}+s_{ij}\Big(s_{i+1,j}+s_{i-1,j}+s_{i,j+1}+s_{i,j-1}\Big),
\end{equation*}
while after flip,
\begin{equation*}
E(\mathfrak{q'})=E_{0}-s_{ij}\Big(s_{i+1,j}+s_{i-1,j}+s_{i,j+1}+s_{i,j-1}\Big),
\end{equation*}
where $E_{0}$ is the energy contribution of all the spins excluding $s_{ij}$ and its four nearest neighbors. As a result, the energy difference is given by
\begin{equation*}
\Delta E=E(\mathfrak{q'})-E(\mathfrak{q})=-2s_{ij}\Big(s_{i+1,j}+s_{i-1,j}+s_{i,j+1}+s_{i,j-1}\Big).
\end{equation*}
An alternative method (as in \texttt{code5.py}) uses the fact that this expression can also be written as 
\begin{equation}
\label{eq:etab}
\Delta E\in \{-8,-4,0,4,8\},
\end{equation}
therefore, one can easily compute $e^{-\beta\Delta E}$ for this set, store them in a five-element array, and then call them according to case.
\subsection{Periodic boundary conditions}
\begin{figure}[H]
	\centering
	\fbox{\rule[0cm]{0cm}{0cm}     \rule[0cm]{0cm}{0cm}
		\includegraphics[scale=0.58]{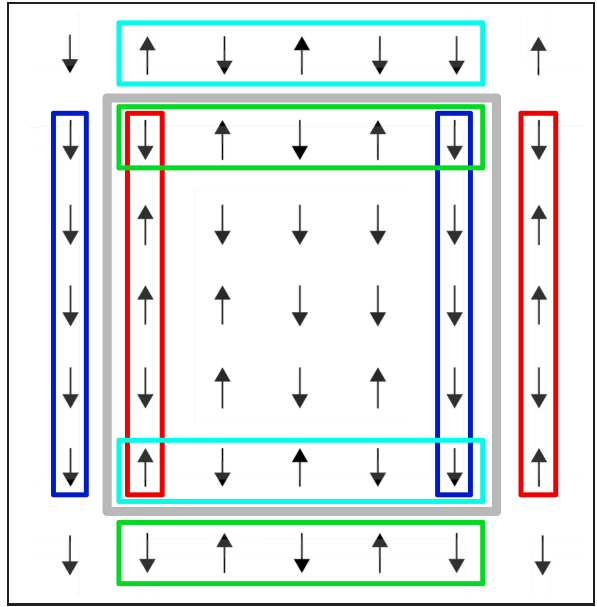}}
	\caption{\label{fig:2}
	The original system of Fig.~\ref{fig:1} demarcated by the gray square, and its four nearest-neighbor rows and columns of spins separated off from the four neighboring image cells. Rectangles with identical colors are evidently of the same spin patterns.}
\end{figure}

Hamiltonians involving many-body interactions are appeared in almost all physics, raising a quite important question as to how to treat the particles lying at the boundary of the system. As an illustration, these boundary particles, compared to the rest lying within the volume of the system, would accordingly experience only half of the interactions involved if the system is entirely immersed in say vacuum. In the thermodynamic limit (i.e., in macroscopic systems) the number $N_b$ of these boundary particles is absolutely negligible compared to the number $N_v$ of particles lying within the volume, therefore they do not affect the results. Nevertheless, the ratio $N_{b}\big/N_{v}$ is quite considerable (this is sometimes referred to as the finite-size effect) in all computer simulations of systems of particles due to the fact that simulating only small-scale systems several orders of magnitude smaller than their macroscopic analogues, as a consequence of limited computational power, is indeed feasible. A clear illustration of such systems has previously been shown in Fig.~\ref{fig:1} in that most of the particles lay on the border. The question then arises: how is it possible to derive, or at least estimate approximately, different properties of interest pertaining to a given macroscopic system by only simulating a microscopic part of it? To do so, also to minimize the finite-size effect, a widely-used computational trick is called periodic boundary conditions---other types of boundary conditions are also possible, such as antiperiodic, free, etc. As an illustration, it is quite routine in computational solid state physics to calculate electronic structures of nanoscale materials~\cite{m1,m2,m3,m4} using only their unit-cell information and then by applying the Born--von Karman periodic boundary conditions~\cite{ash}. In the Ising model, as in MD simulations~\cite{m0,m9,m10}, it is assumed that the original system (the primary cell) is periodically replicated along the two spacial directions (or in the three directions for 3D problems) to form a macroscopic lattice~\cite{haile}. Since only nearest-neighbor interactions are included in the Hamiltonian of the model (Eq.~\ref{eq:h3}), the macroscopic lattice generated by replication, also involving an infinitely large number of particles, is simply scaled down to the original system plus its four nearest-neighbor rows and columns of spins, as illustrated in Fig.~\ref{fig:2}. Such a computational technique considerably reduces the finite-size effect particularly as the number of particles of the system increases.
\subsection{\label{sec:34}Computer codes}
We present five computer codes---\texttt{code1.py}, \texttt{code2.f90}, \texttt{code3.py}, \texttt{code4.f90}, and \texttt{code5.py}---written in both Python~\cite{py} (with \texttt{.py} suffix) and Fortran~\cite{fort} (with \texttt{.f90} suffix). The first two codes simulate a 2D problem, the next two solve a 3D problem, and the last one simulates a 2D lattice in that the energy difference is calculated according to Eq.~\ref{eq:etab}. To run the codes, say \texttt{code1.py}, copy it into a text editor, save it with the same name and suffix, and run in the Linux command-line interface (CLI): \texttt{python3 code1.py}. To run the Fortran codes (say \texttt{code2.f90}), do the same and run: \texttt{gfortran code2.f90 \&\& ./a.out}---one may also replace \texttt{gfortran} with \texttt{ifort}, or \texttt{mpif90}, or any other Fortran compiler. After the run is finished, the file \texttt{output.txt} is generated which contains six columns including, in order from left to right, the inverse temperature $\big(\beta=1\big/k_{B}T\big)$, temperature $(T)$, mean energy $\big(\langle E\rangle\big/J\big)$, mean magnetization $\big(\langle M\rangle\big)$, specific heat $\big(\langle C_{v}\rangle\big/k_{B}\big)$, and zero-field magnetic susceptibility \big($\langle \chi\rangle$ per unit volume\big)---$\langle ...\rangle$ means average over both the total number of spins and number of nearest-neighbors (4 for 2D and 6 for 3D problems). All the codes were written using a function-based programming style (in contrast to the simple-minded top-down design). The temperature range $([0.5,10])$ has been chosen so as to include the analytic critical temperature of the 2D system, namely $k_{B}T_{c}\big/J=2\big/\tanh^{-1}(1/\sqrt{2})=2.269185$, which is an asymptotic value and is approached only in the limit $N\longrightarrow\infty$. As a result, we have used $T_{c}=2.50$ for a $20\times 20$ lattice, obtained by trial and error, for plotting the output of \texttt{code2.f90}. Comparing the simulation times of one given system obtained by both Fortran and Python codes also reveals the fact that the former programming language is much more faster than the latter in scientific computing and simulation.
\subsubsection{code1.py}
\begin{lstlisting}[tabsize=2]
#!/usr/bin/python3
import random
import numpy as np
from numpy.random import rand
#-------------------------------------------
L = 5           # Size of the square lattice  
mcstp = 100     # Number of MC sweep
eqstp = 100     # Number of MC sweep for equilibration
D = 2           # Lattice dimension
N = L**D        # Number of spins
NN = 4          # Number of nearest-neighbors
T_0 = 0.5       # Initial temperature
T_f = 10.0      # Final temperature
dT = 0.1        # Temperature step
#----------------------------------
c = 0           # Counter
T = T_0         # Temperature initialization
norm = 1./(mcstp*N*NN)
nr2 = norm/mcstp
#---------------
def INIT(L):
	""" Generates initial confuguration (cfg): all spins up  """
	cfg = np.random.randint(1, size = (L, L)) + 1
	return cfg
#--------------------------
def METROPOLIS(spin, beta):
	"""The Metropolis algorithm"""
	for i in range(L): 
		for j in range(L):
			x = np.random.randint(0, L) # X coordinate
			y = np.random.randint(0, L) # Y coordinate
			s = spin[x, y]
			# Periodic boundary conditions --------------
			R = spin[(x + 1)%L, y] + spin[x, (y + 1)%L] \
			+ spin[(x - 1)%L, y] + spin[x, (y - 1)%L]
			#--------------------------------------------
			dE = 2*s*R # Energy difference
			if dE < 0.:
				s *= -1 # Flips the spin
			elif rand() < np.exp(-dE*beta):	# Throws the die
				s *= -1 # Flips the spin
			spin[x, y] = s # Returns the spin without flipping
	return spin
#--------------------------
def CALCULATE_ENERGY(spin):
	"""Computes energy"""
	En = 0. # Energy
	for x in range(len(spin)):
		for y in range(len(spin)):
			S = spin[x, y]
			# Periodic boundary conditions --------------
			R = spin[(x + 1)%L, y] + spin[x, (y + 1)%L] \
				+ spin[(x - 1)%L, y] + spin[x, (y - 1)%L]
			#--------------------------------------------			
			En -= S*R
	return En
#---------------------------------
def CALCULATE_MAGNETIZATION(spin):
	"""Computes magnetization"""
	mgnt = np.sum(spin) # Magnetization
	return mgnt
#-----------------------------------
# Opens the output file "output.txt" 
#  containing beta (B= 1/k_{B}T), 
#  temperature (T), and the mean values of energy (E), 
#  magnetization (M), specific heat (C_v), 
#  and susceptibility (Chi) of the system 
#  at every temperature step.
g = open('output.txt', 'w')
g.write('# T^{-1}        T           E           M           C_v         Chi\n')
g.write('#=======    ========     ========    ========    ========    ========\n')
#--------------------------------------------------
while T <= T_f:		# Temperature loop (main loop)
	T += dT
	summ = [0, 0, 0, 0]	# Accumulators
	spin = INIT(L)
	c = c + 1
	B = 1./T; B2 = B*B
	for sweep in range(eqstp):  # Equilibration loop
		METROPOLIS(spin, B)
#---------------------------------------------
	for sweep in range(mcstp):  # Main MC loop
		METROPOLIS(spin, B)
		E = CALCULATE_ENERGY(spin)
		M = CALCULATE_MAGNETIZATION(spin)
		#---------------------------------
		summ[0] += E      # E accumulator
		summ[1] += E*E    # E^2 accumulator
		summ[2] += M      # M accumulator
		summ[3] += M*M    # M^2 accumulator
#-------------------------------------------------
# Thermodynamic averages at each temperature step
	mean_E = summ[0]*norm  # Mean E
	mean_M = summ[2]*norm  # Mean M
	C_v = (norm*summ[1] - nr2*summ[0]*summ[0])*B2  # Specific heat (C_v)
	chi = (norm*summ[3] - nr2*summ[2]*summ[2])*B   # Susceptibility (Chi)
	g.write('%f    %f    %f    %f    %f    %f\n' \
	%(B, T, mean_E, mean_M, C_v, chi))
	print('>>> Temperature step %d' %(c))
\end{lstlisting}
\subsubsection{code2.f90}
\begin{lstlisting}[tabsize=2]
module SHARED_DATA
implicit none
save
integer i, j	! Loop counters
integer x, y	! Position coordinates
integer up, down, right, left
integer, parameter :: L = 5         ! Size of the quare lattice
integer, parameter :: D = 2         ! Lattice dimension
integer, parameter :: N = L**D      ! Number of spins
integer, parameter :: mcstp = 100   ! Monte Carlo sweeps
integer, parameter :: eqstp = 100   ! Monte Carlo sweeps for equilibration
integer, parameter :: NN = 4        ! Number of nearest-neighbors
real, parameter :: T_0 = 0.5        ! Initial temperature
real, parameter :: T_f = 10.0       ! Final temperature
real, parameter :: norm = 1./(mcstp*N*NN)
real, parameter :: nr2 = norm/mcstp
real, parameter :: dT = 0.1         ! Temperature step
end module
!-------------------------------
subroutine METROPOLIS(sp, beta)
! The Metropolis algorithm"""
use SHARED_DATA
implicit none
integer, intent(inout) :: sp(L, L)	! Spin
real, intent(in) :: beta	! = 1/k_{B}T
integer s	! Box for spin
real R
integer dE	! Energy difference
real rand, x0, y0	! Random numbers
do i = 1, L 
	do j = 1, L 
		call random_number(x0)
		call random_number(y0)
		x = int(1 + (L - 1)*x0)
		y = int(1 + (L - 1)*y0)
		s = sp(x, y)
		! Periodic boundary conditions
		if (x == 1) then
			left = sp(L, y)
			right = sp(2, y)
		else if (x == L) then
			left = sp(L - 1, y)
			right = sp(1, y)
		else
			left = sp(x - 1, y)
			right = sp(x + 1, y)
		end if
		if (y == 1) then
			up = sp(x, 2)
			down = sp(x, L)
		else if (y == L) then
			up = sp(x, 1)
			down = sp(x, L - 1)
		else
			up = sp(x, y + 1)
			down = sp(x, y - 1)
		end if ! End of periodic boundary conditions
		R = up + down + right + left
		dE = 2*s*R
		call random_number(rand)
		if (dE < 0.) then
			s = -s  ! Flips the spin
		else if (rand < exp(-dE*beta)) then	! Throws the die
			s = -s  ! Flips the spin
		end if
		sp(x, y) = s  ! Returns the spin without flipping
	enddo
enddo
end subroutine METROPOLIS
!-----------------------------------
real function CALCULATE_ENERGY(sp1)
use SHARED_DATA
implicit none
integer, intent(in) :: sp1(L, L)	! Spin
real En, R
En = 0.	! Energy initialization
do x = 1, L
	do y = 1, L 
		if (x == 1) then
			left = sp1(L, y)
			right = sp1(2, y)
		else if (x == L) then
			left = sp1(L - 1, y)
			right = sp1(1, y)
		else
			left = sp1(x - 1, y)
			right = sp1(x + 1, y)
		end if
		if (y == 1) then
			up = sp1(x, 2)
			down = sp1(x, L)
		else if (y == L) then
			up = sp1(x, 1)
			down = sp1(x, L - 1)
		else
			up = sp1(x, y + 1)
			down = sp1(x, y - 1)
		end if
		R = up + down + right + left
		En = En - sp1(x, y)*R
	enddo
enddo
CALCULATE_ENERGY = En	! Energy
end function
!------------------------------------------
real function CALCULATE_MAGNETIZATION(sp2)
use SHARED_DATA
implicit none
integer, intent(in) :: sp2(L, L)	! Spin
CALCULATE_MAGNETIZATION = sum(sp2)	! Magnetization
end function
!------------------------------
program ising	! Main program
use SHARED_DATA
implicit none
integer sweep	! Loop counter
real T, B, B2	! Temperature, beta = 1/k_{B}T, beta^2
real summ0, summ1, summ2, summ3, summ4	! Accumulators
real E, M ! energy, magnetization
real CALCULATE_ENERGY, CALCULATE_MAGNETIZATION ! External functions
real mean_E, mean_M, C_v, chi ! Mean E, mean M, specific heat, susceptibility
integer :: spin(L, L) = 1 ! Initial configuration: all spins up
integer c	! Counter
open(1, file = 'output.txt') ! Output file: beta, T, E, M, Cv, chi
open(2, file = 'eqltst.txt') ! Output file for equilibration test
T = T_0		! Temperature initialization
c = 0
10 format (6f12.6)
write(1, *)"#   T^{-1}         T           E          M          C_v         Chi"
write(1, *)"#  ========    ========    ========    ========    ========    ========"
do while (T <= T_f)		! Temperature loop (main loop)
	T = T + dT
	c = c + 1
	summ0 = 0.; summ1 = 0.; summ2 = 0.; summ3 = 0. ! Initializing accumulators
	B = 1./T; B2 = B*B ! Beta, beta^2
	do sweep = 1, eqstp ! Equilibration loop
		call METROPOLIS(spin, B)
		! Activate the following 4 comment lines
		! (from the line 142 to 145) to check whether 
		! equilibration is achieved for a given 
		! temperature (here, T = 2). 
		! To this end, plot eqltst.txt.
		!E = CALCULATE_ENERGY(spin)
		!if (int(T) == 2) then
		!	write(2,*) step, E
		!end if
	enddo
	!-------------------------------------
	do sweep = 1, mcstp     ! Main MC loop 
		call METROPOLIS(spin, B)
		E = CALCULATE_ENERGY(spin)
		M = CALCULATE_MAGNETIZATION(spin)
		summ0 = summ0 + E       ! E accumulator
		summ1 = summ1 + E*E     ! E^2 accumulator
		summ2 = summ2 + M       ! M accumulator
		summ3 = summ3 + M*M     ! M^2 accumulator
	end do
	!--------------------------------------------------
	! Thermodynamic averages at each temperature value
	mean_E = summ0*norm       ! Mean energy
	mean_M = summ2*norm       ! Mean magnetization 
	C_v = (norm*summ1 - nr2*summ0*summ0)*B2    ! Specific heat
	chi = (norm*summ3 - nr2*summ2*summ2)*B     ! Susceptibility
	write(1, 10)B, T, mean_E, mean_M, C_v, chi
	write(*, *)" >>> Temperature step", c
end do
end program
\end{lstlisting}
\subsubsection{code3.py}
\begin{lstlisting}[tabsize=2]
#!/usr/bin/python3
import random
import numpy as np
from numpy.random import rand
#--------------------------------------------
L = 4           # Size of the square lattice 
D = 3           # Lattice dimension 
mcstp = 100     # Number of MC sweeps
eqstp = 100     # Number of MC sweeps for equilibration
N = L**D        # Number of spins
NN = 6          # Number of nearest-neighbors
T_0 = 0.5       # Initial temperature
T_f = 10.0      # Final temperature
dT = 0.1        # Temperature step
c = 0           # Counter
T = T_0         # Temperature initialization
norm = 1./(mcstp*N*NN)
nr2 = norm/mcstp
#---------------
def INIT(L):
	"""Initial confuguration (cfg): all spins up"""
	cfg = np.random.randint(1, size = (L, L, L)) + 1
	return cfg
#--------------------------
def METROPOLIS(spin, beta):
	"""The Metropolis algorithm"""
	for i in range(L): 
		for j in range(L):
			for k in range(L):
				x = np.random.randint(0, L)		# X coordinate
				y = np.random.randint(0, L)		# Y coordinate
				z = np.random.randint(0, L)		# Z coordinate
				s = spin[x, y, z]	# Spin
				# Periodic boundary conditions --------------------
				R = spin[(x + 1)%L, y, z] + spin[x, (y + 1)%L, z] \
				+ spin[(x - 1)%L, y, z] + spin[x, (y - 1)%L, z] \
				+ spin[x, y, (z - 1)%L] + spin[x, y, (z + 1)%L]
				#----------------------------------------------
				dE = 2*s*R		# Energy difference
				if dE < 0.:
					s *= -1		# Flips the spin
				elif rand() < np.exp(-dE*beta):	# Throws the die
					s *= -1		# Flips the spin
				spin[x, y, z] = s	# Returns the spin without flipping
	return spin
#--------------------------
def CALCULATE_ENERGY(spin):
	"""computes energy"""
	En = 0.
	for x in range(len(spin)):
		for y in range(len(spin)):
			for z in range(len(spin)):
				S = spin[x, y, z]	# Spin
				# Periodic boundary conditions --------------------
				R = spin[(x + 1)%L, y, z] + spin[x, (y + 1)%L, z] \
				+ spin[(x - 1)%L, y, z] + spin[x, (y - 1)%L, z] \
				+ spin[x, y, (z - 1)%L] + spin[x, y, (z + 1)%L]
				#----------------------------------------------			
				En -= S*R	# Energy
	return En
#----------------------------------
def CALCULATE_MAGNETIZATION(spin):
	"""computes magnetization"""
	mgnt = np.sum(spin)		# Magnetization
	return mgnt
#-----------------------------------
# Opens the output file "output.txt" 
#  containing beta (B= 1/k_{B}T), 
#  temperature (T), and the mean values of energy (E), 
#  magnetization (M), specific heat (C_v), 
#  and susceptibility (Chi) of the system 
#  at every temperature.
g = open('output.txt', 'w')
g.write('# T^{-1}        T           E           M           C_v         Chi\n')
g.write('#=======    ========     ========    ========    ========    ========\n')
#-----------------------------------------------
while T <= T_f:  # Temperature loop (main loop)
	T += dT
	s = [0, 0, 0, 0]	# Initializing accumulators
	spin = INIT(L)		# Loads initial configuration
	c = c + 1
	B = 1./T; B2 = B*B
	for sweep in range(eqstp):	# Equilibration loop
		METROPOLIS(spin, B)
	#------------------------------------------
	for sweep in range(mcstp):	# Main MC loop 
		METROPOLIS(spin, B)
		E = CALCULATE_ENERGY(spin)
		M = CALCULATE_MAGNETIZATION(spin)
		#---------------------------------
		s[0] += E		# E accumulator
		s[1] += E*E		# E^2 accumulator
		s[2] += M		# M accumulator
		s[3] += M*M		# M^2 accumulator
		#---------------------------------------------
	# Thermodynamic averages at each temperature step
	mean_E = s[0]*norm  # Mean energy
	mean_M = s[2]*norm  # Mean magnetization 
	C_v = (norm*s[1] - nr2*s[0]*s[0])*B2 # Specific heat
	chi = (norm*s[3] - nr2*s[2]*s[2])*B   # Susceptibility
	g.write('%f    %f    %f    %f    %f    %f\n' \
	%(B, T, mean_E, mean_M, C_v, chi))
	print('>>> Temperature step %d' %(c))
\end{lstlisting}
\subsubsection{code4.f90}
\begin{lstlisting}[tabsize=2]
module shared_data
implicit none
save
integer i, j, k ! Loop counters
integer x, y, z ! Spins' position coordinates
integer up, down, right, left, above, below
integer, parameter :: L = 4	        ! Size of the square lattice 
integer, parameter :: D = 3         ! Lattice dimension
integer, parameter :: N = L**D      ! Number of spins
integer, parameter :: NN = 6        ! Number of nearest-neighbors
integer, parameter :: mcstp = 100   ! Number of Monte Carlo sweeps 
integer, parameter :: eqstp = 100   ! Number of Monte Carlo sweeps for equilibration
real, parameter :: T_0 = 0.5        ! Initial temperature
real, parameter :: T_f = 10.0       ! Final temperature
real, parameter :: norm = 1./(mcstp*N*NN)
real, parameter :: nr2 = norm/mcstp
real, parameter :: dT = 0.1         ! Temperature step
end module
!--------------------------------
subroutine METROPOLIS(spin, beta)
! The Metropolis algorithm
use shared_data
implicit none
integer, intent(inout) :: spin(L, L, L)	! Spins
real, intent(in) :: beta	! Beta = 1/kB*T
integer s	! Box for spin
real R
real dE		! Energy difference
real x0, y0, z0, rand	! Random numbers
do i = 1, L 
	do j = 1, L 
		do k = 1, L
			call random_number(x0)
			call random_number(y0)
			call random_number(z0)
			x = int(1 + (L - 1)*x0)		! X coordinate
			y = int(1 + (L - 1)*y0)		! Y coordinate
			z = int(1 + (L - 1)*z0)		! Z coordinate
			s = spin(x, y, z)	! Spin
			! Periodic boundary conditions
			if (x == 1) then
				left = spin(L, y, z)
				right = spin(2, y, z)
			else if (x == L) then
				left = spin(L - 1, y, z)
				right = spin(1, y, z)
			else
				left = spin(x - 1, y, z)
				right = spin(x + 1, y, z)
			end if
			if (y == 1) then
				up = spin(x, 2, z)
				down = spin(x, L, z)
			else if (y == L) then
				up = spin(x, 1, z)
				down = spin(x, L - 1, z)
			else
				up = spin(x, y + 1, z)
				down = spin(x, y - 1, z)
			end if
			if (z == 1) then
				above = spin(x, y, 2)
				below = spin(x, y, L)
			else if (z == L) then
				above = spin(x, y, 1)
				below = spin(x, y, L - 1)
			else
				above = spin(x, y, z + 1)
				below = spin(x, y, z - 1)
			end if 		! End of periodic boundary conditions
			R = up + down + right + left + above + below
			dE = 2*s*R		! Energy difference
			call random_number(rand)	!Throws the die
			if (dE < 0.) then
				s = -s		! Flips the spin
			elseif (rand < exp(-dE*beta)) then
				s = -s		! Flips the spin
			endif
			spin(x, y, z) = s	! Returns the spin without flipping
		enddo
	enddo
enddo
end subroutine METROPOLIS
!-----------------------------------
real function CALCULATE_ENERGY(spin)
use shared_data
implicit none
integer, intent(in) :: spin(L, L, L)
real En, R
En = 0. ! Energy initialization
do x = 1, L
	do y = 1, L 
		do z = 1, L
			! Periodic boundary conditions
			if (x == 1) then
				left = spin(l, y, z)
				right = spin(2, y, z)
			else if (x == L) then
				left = spin(l - 1, y, z)
				right = spin(1, y, z)
			else
				left = spin(x - 1, y, z)
				right = spin(x + 1, y, z)
			end if
			if (y == 1) then
				up = spin(x, 2, z)
				down = spin(x, L, z)
			else if (y == L) then
				up = spin(x, 1, z)
				down = spin(x, L - 1, z)
			else
				up = spin(x, y + 1, z)
				down = spin(x, y - 1, z)
			end if
			if (z == 1) then
				above = spin(x, y, 2)
				below = spin(x, y, L)
			else if (z == L) then
				above = spin(x, y, 1)
				below = spin(x, y, L - 1)
			else
				above = spin(x, y, z + 1)
				below = spin(x, y, z - 1)
			end if
			R = up + down + right + left + above + below
			En = En - spin(x, y, z)*R	! Energy accumulation
		enddo
	enddo
enddo
CALCULATE_ENERGY = En	! Energy
end function
!------------------------------------------
real function CALCULATE_MAGNETIZATION(spin)
use shared_data
implicit none
integer, intent(in) :: spin(L, L, L)	! Spin
CALCULATE_MAGNETIZATION = sum(spin)	! Magnetization
end function
!------------------------------
program ising	! Main program
use shared_data
implicit none
integer sweep	! Loop counter
real T, B, B2	! Temperature, beta = 1/k_{B}T, beta^2
real summ0, summ1, summ2, summ3, summ4	! Accumulators
real E, M	! Energy, magnetization
real CALCULATE_ENERGY, CALCULATE_MAGNETIZATION	! external functions
real mean_E, mean_M, C_v, chi	! Mean E, mean M, specific heat, susceptibility
integer :: spin(L, L, L) = 1 ! Initial configuration: all spins up
integer c	! Counter
open(1, file = 'output.txt')	! Output file
!--------------------------------------------
T = T_0		! Temperature initialization
c = 0
10 format (6f12.6)
write(1, *)"#   T^{-1}         T           E          M          C_v         Chi"
write(1, *)"#  ========    ========    ========    ========    ========    ========"
do while (T <= T_f)		! Temperature loop (main loop)
	T = T + dT
	c = c + 1
	! Initializing accumulators ------------------
	summ0 = 0.; summ1 = 0.; summ2 = 0.; summ3 = 0.
	!---------------------------------
	B = 1./T; B2 = B*B	! Beta, beta^2
	do sweep = 1, eqstp	! Equilibration loop
		call METROPOLIS(spin, B)
	enddo
	!---------------------------------
	do sweep = 1, mcstp ! Main MC loop 
		call METROPOLIS(spin, B)
		E = CALCULATE_ENERGY(spin)
		M = CALCULATE_MAGNETIZATION(spin)
		summ0 = summ0 + E      ! E accumulator
		summ1 = summ1 + E*E    ! E^2 accumulator
		summ2 = summ2 + M      ! M accumulator
		summ3 = summ3 + M*M    ! M^2 accumulator
	enddo
	!--------------------------------------------------
	! Thermodynamic averages at each temperature value
	mean_E = summ0*norm		! Mean energy
	mean_M = summ2*norm		! Mean magnetization 
	C_v = (norm*summ1 - nr2*summ0*summ0)*B2		! Specific heat
	chi = (norm*summ3 - nr2*summ2*summ2)*B		! Susceptibility
	write(1, 10)B, T, mean_E, mean_M, C_v, chi
	write(1, 10)B, T, mean_E, mean_M, C_v, chi
	write(*, *)" >>> Temperature step", c
enddo
end program
\end{lstlisting}
\subsubsection{code5.py}
\begin{lstlisting}[tabsize=2]
#!/usr/bin/python3
import random
import numpy as np
from numpy.random import rand
#-------------------------------------------
L = 5           # Size of the square lattice  
mcstp = 1000    # Number of MC sweeps
eqstp = 100     # Number of MC sweeps for equilibration
D = 2           # Lattice dimension
N = L**D        # Number of spins
NN = 4          # Number of nearest-neighbors
T = 2.5         # Temperature
B = 1./T        # Beta = 1/k_{B}T
B2 = B*B        # Beta^2
norm = 1./(mcstp*N*NN)
nr2 = norm/mcstp
#----------------
def INIT(L):
	""" Generates initial confuguration (cfg): all spins up  """
	cfg = np.random.randint(1, size = (L, L)) + 1
	return cfg
#--------------
def bw(dE):
	'''Computes Boltzmann weights'''
	if dE == -8 :
		return 24.532530197     # exp(-dE/T)=exp(8/2.5)
	elif dE == -4 :
		return 4.953032424      # exp(-dE/T)=exp(4/2.5)
	elif dE == 0 :
		return 1.0              # exp(-dE/T)=exp(0)
	elif dE == 4 :
		return 0.201896518      # exp(-dE/T)=exp(-4/2.5)
	elif dE == 8 :
		return 0.040762204      # exp(-dE/T)=exp(-8/2.5)
#-------------------------------------------------------
def METROPOLIS(spin, beta):
	"""The Metropolis algorithm"""
	for i in range(L): 
		for j in range(L):
			x = np.random.randint(0, L) # X coordinate
			y = np.random.randint(0, L) # Y coordinate
			s = spin[x, y]
			# Periodic boundary conditions --------------
			R = spin[(x + 1)%L, y] + spin[x, (y + 1)%L] \
			+ spin[(x - 1)%L, y] + spin[x, (y - 1)%L]
			#--------------------------------------------
			dE = int(2*s*R) # Energy difference
			if dE < 0.:
				s *= -1 # Flips the spin
			elif rand() < bw(dE):	# Throws the die
				s *= -1 # Flips the spin
			spin[x, y] = s # Returns the spin without flipping
	return spin
#--------------------------
def CALCULATE_ENERGY(spin):
	"""Computes energy"""
	En = 0. # Energy
	for x in range(len(spin)):
		for y in range(len(spin)):
			S = spin[x, y]
			# Periodic boundary conditions --------------
			R = spin[(x + 1)%L, y] + spin[x, (y + 1)%L] \
				+ spin[(x - 1)%L, y] + spin[x, (y - 1)%L]
			#--------------------------------------------			
			En -= S*R
	return En
#---------------------------------
def CALCULATE_MAGNETIZATION(spin):
	"""Computes magnetization"""
	mgnt = np.sum(spin) # Magnetization
	return mgnt
#--------------------------------------
summ = [0, 0, 0, 0, 0]	# Accumulators
spin = INIT(L)
B = 1./T; B2 = B*B
# Equilibration loop -----
for sweep in range(eqstp):  
	METROPOLIS(spin, B)
#------------------------------------------------------
# Storing initial configuration in the file "spin.xyz"
# The file will contain all spins at every sweep,
# and will be used as the input file for XMakemol.
f = open('spin.xyz', 'w')
f.write("%d\n\n" %(N))
for x in range(L):
	for y in range(L):
		if spin[x, y] == 1:
			f.write("O %d %d\n" %(x, y))
               # O means oxygen: XMakemol parameter
		elif spin[x, y] == -1:
			f.write("Na %d %d\n" %(x, y))
               # Na means sodium: XMakemol parameter
#----------------------------------------------------
for sweep in range(mcstp):  # Main MC loop
	METROPOLIS(spin, B)
	E = CALCULATE_ENERGY(spin)
	M = CALCULATE_MAGNETIZATION(spin)
	#---------------------------------
	summ[0] += E      # E accumulator
	summ[1] += E*E    # E^2 accumulator
	summ[2] += M      # M accumulator
	summ[3] += M*M    # M^2 accumulator
	summ[4] += abs(M) # |M| accumulator
	#-----------------------------------
	f = open('spin.xyz', 'a')
	for x in range(L):
		for y in range(L):
			if spin[x, y] == 1:
				f.write("O %d %d\n" %(x, y))
			elif spin[x, y] == -1:
				f.write("Na %d %d\n" %(x, y))
#--------------------------------------------
# Computes thermodynamic averages
mean_E = summ[0]*norm      # <E>
mean_E2 = summ[1]*norm     # <E^2> 
mean_M = summ[2]*norm      # <M> 
mean_M2 = summ[3]*norm     # <M^2> 
mean_abs_M = summ[4]*norm  # <|M|> 
C_v = (mean_E2 - nr2*summ[0]*summ[0])*B2  # Specific heat
chi = (mean_M2 - nr2*summ[2]*summ[2])*B   # Susceptibility
print('\n -----------------------------------------------------------------')
print(' >>> THERMODYNAMIC AVERAGES:\n')
print('                              Lattice size  =  %d*%d' %(L, L))
print('                           Temperature [T]  =  %f' %(T))
print('                         Mean energy [<E>]  =  %f J' %(mean_E))
print('                                     <E^2>  =  %f J^2' %(mean_E2))
print('                   Mean manetization [<M>]  =  %f' %(mean_M))
print('                                     <M^2>  =  %f' %(mean_M2))
print('                                     <|M|>  =  %f' %(mean_abs_M))
print('        Energy fluctuation [<E^2> - <E>^2]  =  %f J^2' %(mean_E2 - nr2*summ[0]*summ[0]))
print(' Magnetization fluctuation [<M^2> - <M>^2]  =  %f' %(mean_M2 - nr2*summ[2]*summ[2]))
print('               Specific heat [<C_v>/k_{B}]  =  %f' %(C_v))
print('                    Susceptibility [<Chi>]  =  %f' %(chi))
print(' -----------------------------------------------------------------\n')
\end{lstlisting}
\subsection{Output of code2.f90}
\begin{figure}[H]
	\centering
	\fbox{\rule[0cm]{0cm}{0cm}     \rule[0cm]{0cm}{0cm}
		\subfigure[]{\label{subfig:3(c)}
			\includegraphics[scale=0.31,angle=-90]{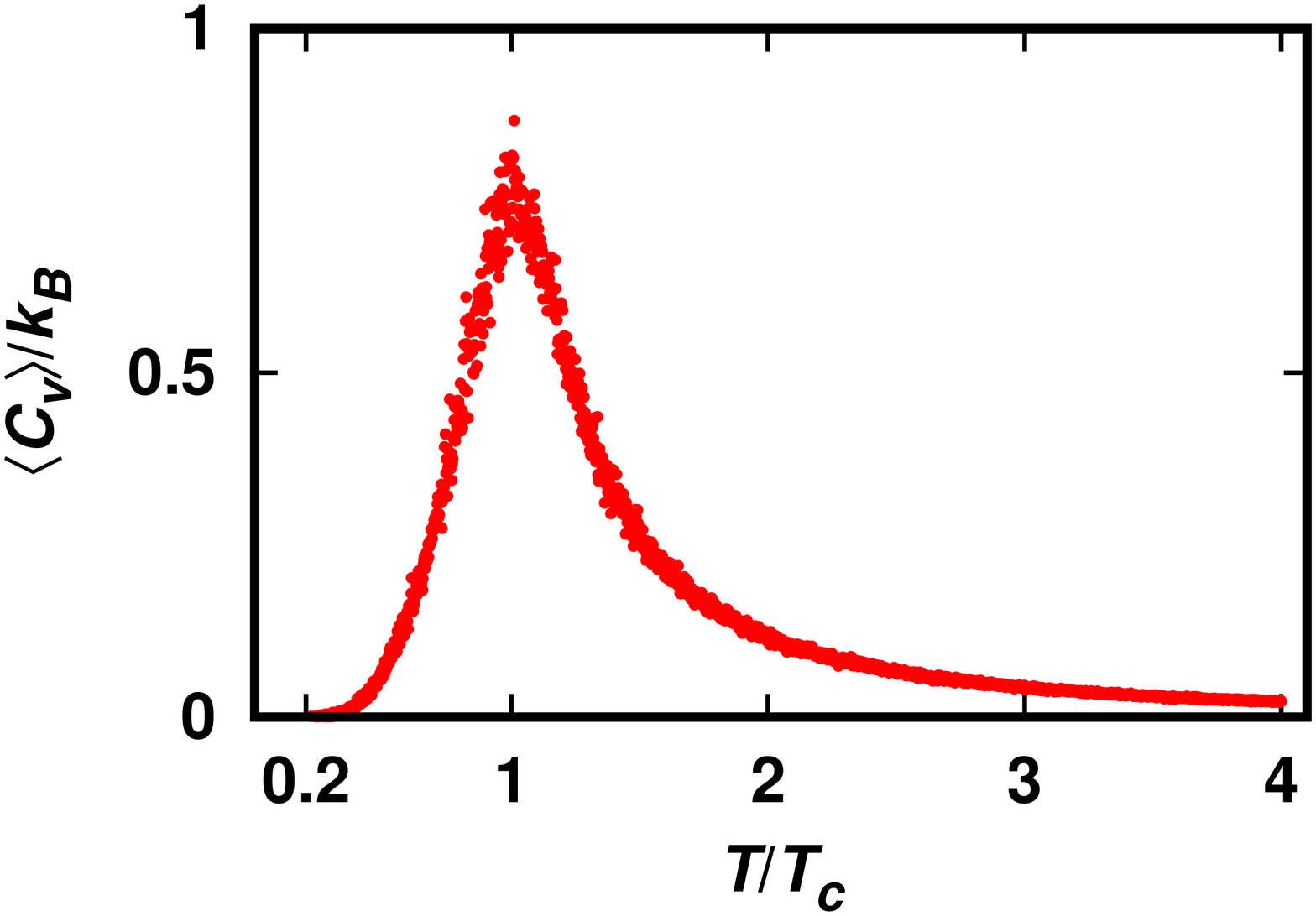}}
		\subfigure[]{\label{subfig:3(d)}
			\includegraphics[scale=0.31,angle=-90]{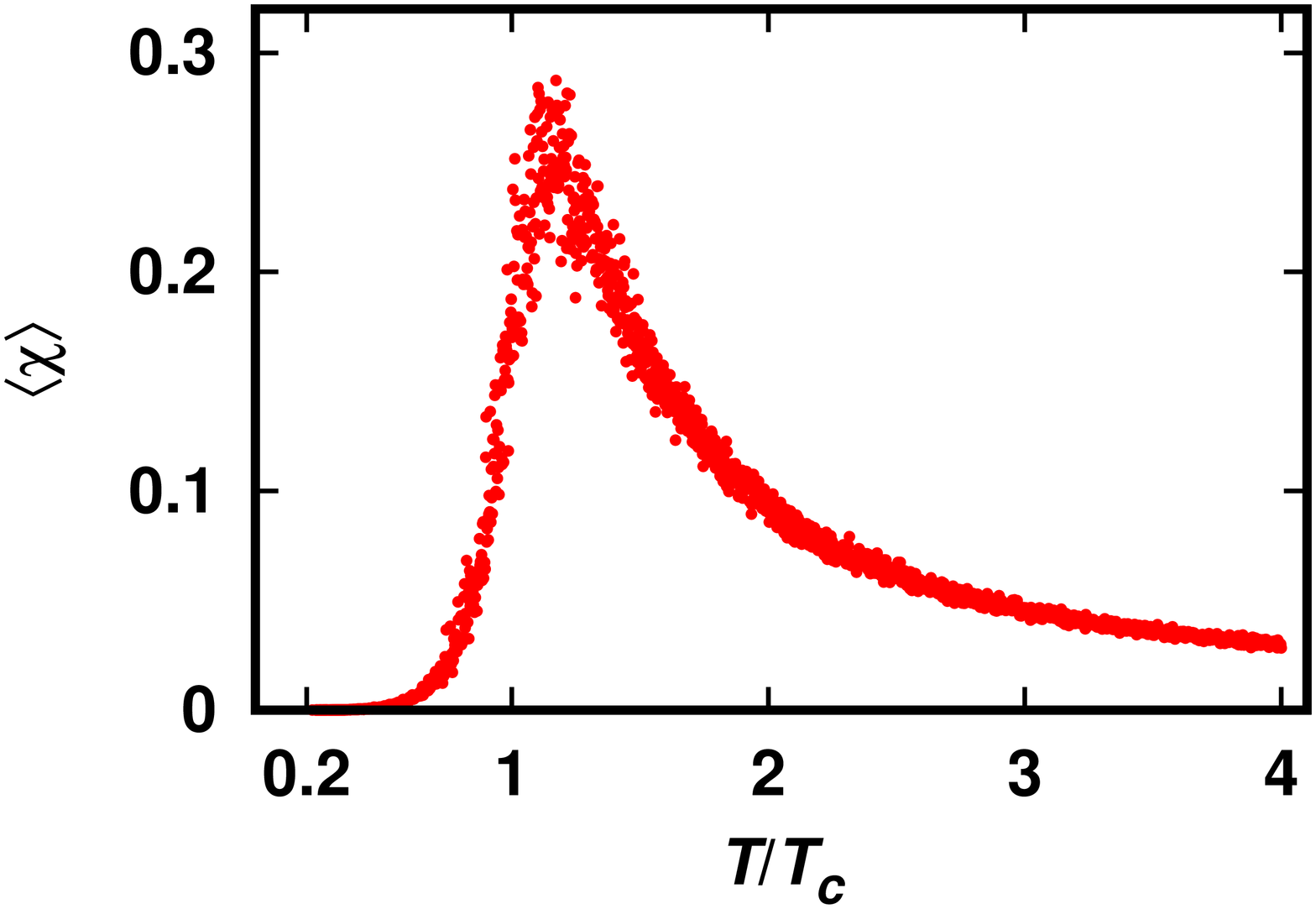}}}
	\caption{\label{fig:3}
		Outputs of \texttt{code2.f90} for a 2D, $20\times 20$, square lattice of spin--\small{1/2} particles, including the temperature dependence of (a) mean energy, (b) mean magnetization, (c) specific heat, and (d) zero-field magnetic susceptibility---rendered in Gnuplot (version 5.2)~\cite{gnu}. The critical temperature is also $T_{c}=2.5$ obtained by trial and error.}
\end{figure}

Fig.~\ref{fig:3} illustrates outputs of \texttt{code2.f90} obtained for a $20\times 20$ lattice with parameters \texttt{eqstp\hspace{0.7mm}=\hspace{0.7mm}mcstp\hspace{0.7mm}=\hspace{0.7mm}2000} and \texttt{dT\hspace{0.7mm}=\hspace{0.7mm}0.005}. 
\begin{figure}[H]
	\centering
	\fbox{\rule[0cm]{0cm}{0cm}     \rule[0cm]{0cm}{0cm}
		\subfigure[]{\label{subfig:3(a)}
			\includegraphics[scale=0.31,angle=-90]{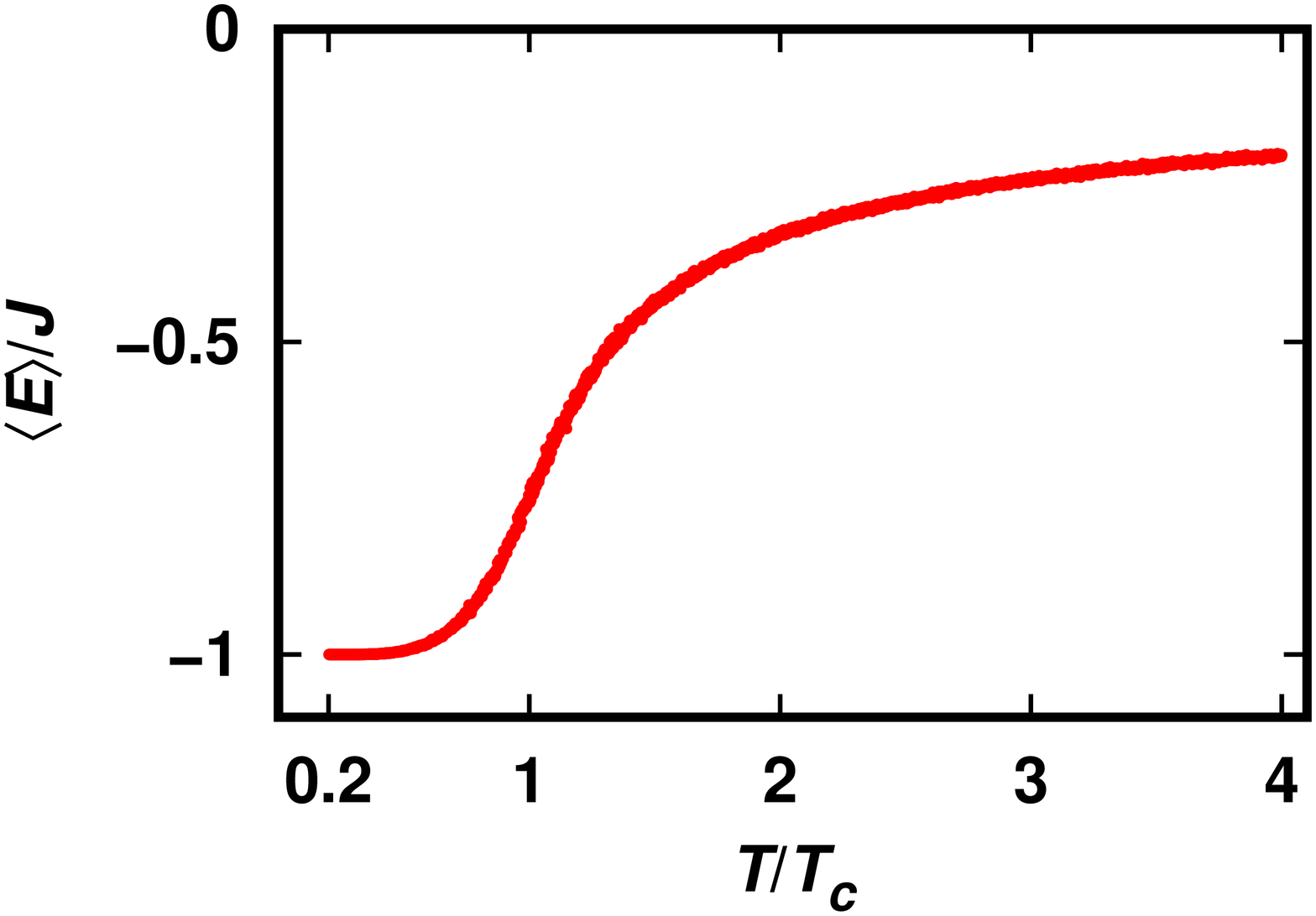}}
		\subfigure[]{\label{subfig:3(b)}
			\includegraphics[scale=0.31,angle=-90]{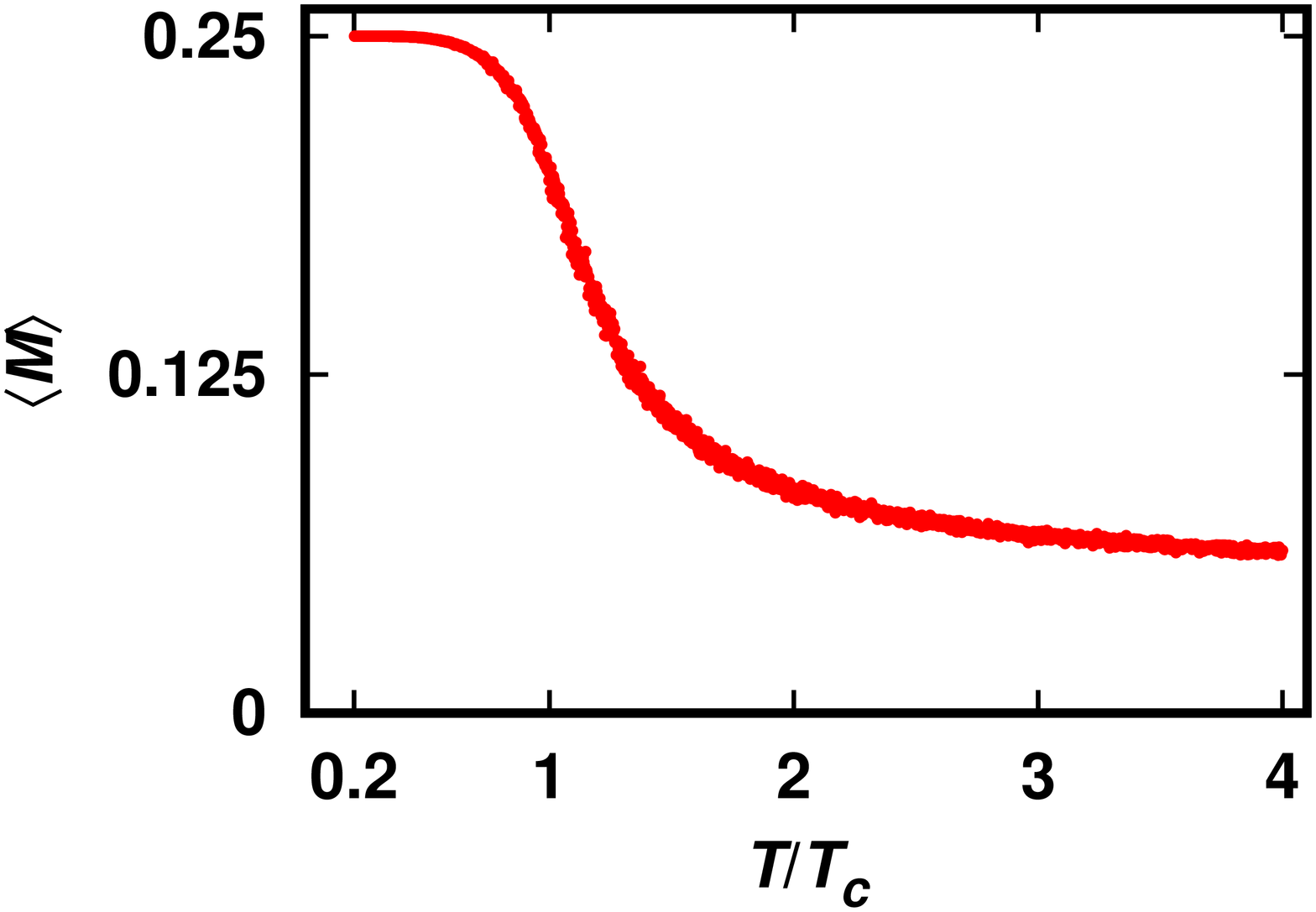}}}
\end{figure}
Phase transition is clearly observed from sharp changes in the diagrams of energy [Fig.~\ref{subfig:3(a)}] and magnetization [Fig.~\ref{subfig:3(b)}], also from the maximums of specific heat [Fig.~\ref{subfig:3(c)}] and zero-field magnetic susceptibility [Fig.~\ref{subfig:3(d)}], taking place at $T_{c}=2.5$. That the energy curve starts with its minimum value ($-J$ at $T\big/T_{c}=0.2$) is due to the fact that the initial configuration of the system has been chosen a minimum-energy microstate (all spins up). At $T=T_c$ energy undergoes a sharp raise, which, according to the thermodynamic relation $C_{v}=(\partial E\big/\partial T)_{N,V}$ gives rise to a maximum at this point. Maximums of the specific heats of solids, such as atomic clusters~\cite{m6}, are also clear manifestations of the solid-liquid phase transition: at the critical temperature, the energy of the system suddenly increases due to the fact that the binding energy of atoms is dominated by thermal fluctuations, increasing as well the kinetic energy of the atoms. The specific heat has been calculated using the energy fluctuation relation of the canonical ensemble
\begin{equation*}
C_v=\frac{\langle(\Delta E)^2\rangle}{k_{B}T^{2}}=\frac{\langle E^2\rangle-\langle E\rangle^2}{k_{B}T^{2}}.
\end{equation*}
In contrast, the temperature dependence of $\langle M\rangle$, starting with its maximum, sharply decreases at the critical temperature and asymptotically approaches zero at high enough temperatures due to increase in the entropy of the system. Such a marked decrease is concurrent with the maximum in the temperature-dependent zero-field magnetic susceptibility of the system, calculated using the fluctuation-susceptibility relation
\begin{equation*}
\chi=\frac{\langle(\Delta M)^2\rangle}{k_{B}T}=\frac{\langle M^2\rangle-\langle M\rangle^2}{k_{B}T}.
\end{equation*}
\subsection{Output of code5.py: Visualizing the Ising lattice using XMakemol}
\begin{figure}[H]
	\centering
	\fbox{\rule[0cm]{0cm}{0cm}     \rule[0cm]{0cm}{0cm}
		\subfigure[sweep 1]{
			\includegraphics[scale=0.0734]{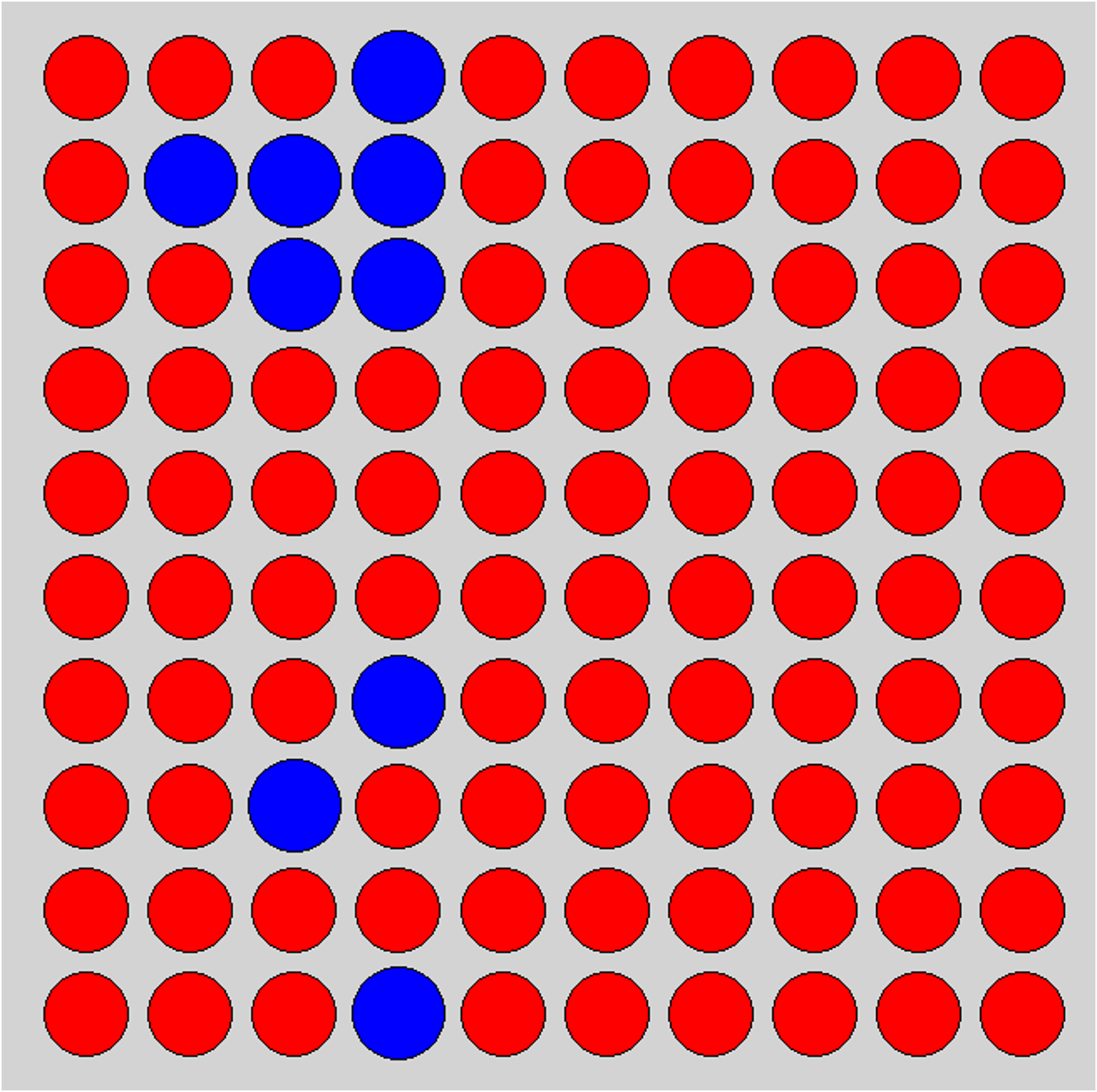}}
		\subfigure[sweep 10]{
			\includegraphics[scale=0.0734]{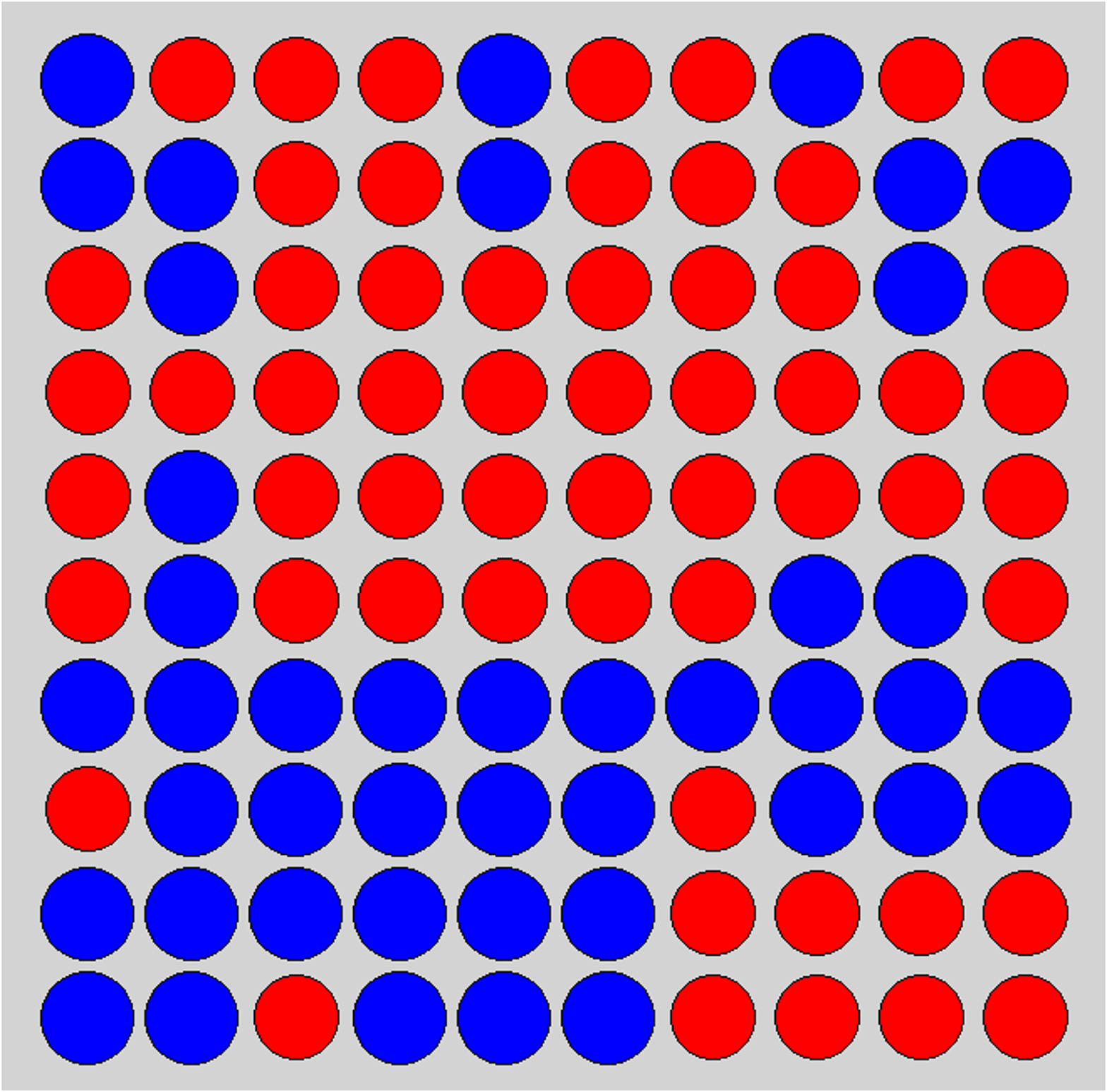}}
				\subfigure[sweep 850]{
			\includegraphics[scale=0.0734]{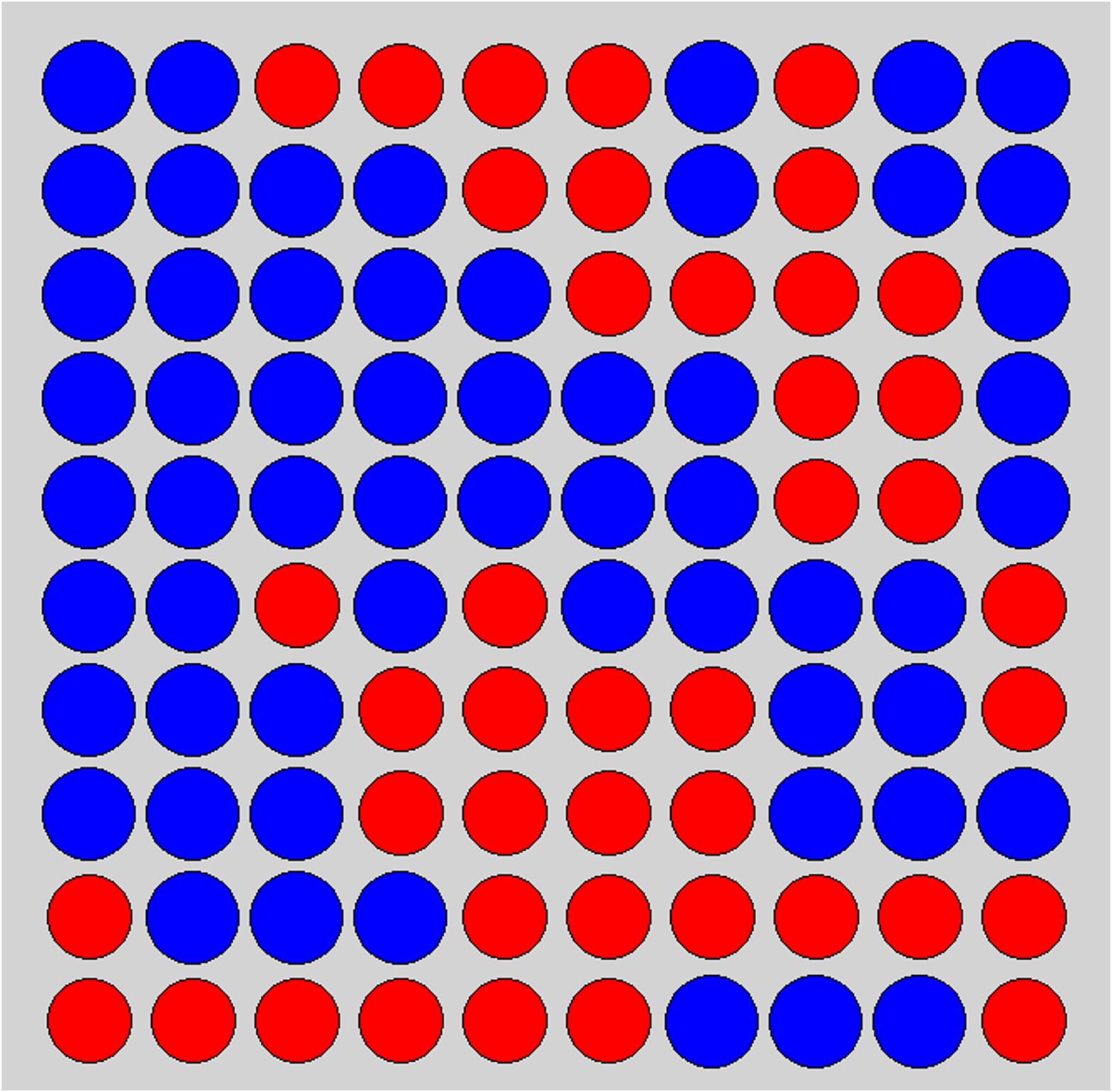}}}
	\caption{\label{fig:4}
	A visualization of the outputs of \texttt{code5.py} simulating a 2D, $10\times 10$, square lattice of spin--\small{1/2} particles at $T=2.5$ at three steps including (a) 1, (b) 10, and (c) 850. The red and blue balls respectively denote the up and down spins.}
\end{figure}

\begin{figure}[H]
	\centering
	\fbox{\rule[0cm]{0cm}{0cm}     \rule[0cm]{0cm}{0cm}
		\includegraphics[scale=0.113]{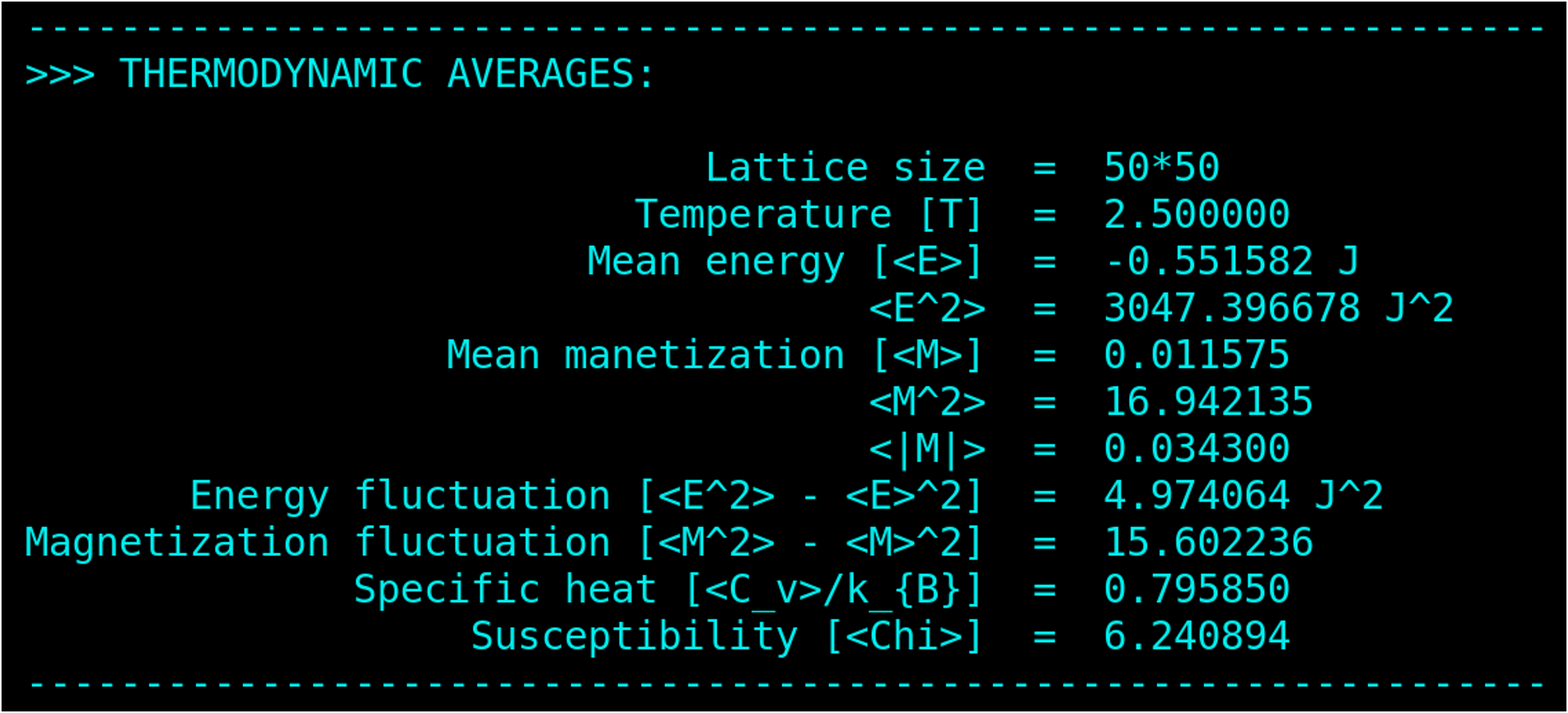}}
	\caption{\label{fig:5}
		Output of \texttt{code5.py} simulating a 2D, $50\times 50$, square lattice of spin--\small{1/2} particles at $T=2.5$.}
\end{figure}

Fig.~\ref{fig:4} illustrates the output of \texttt{code5.py} for a $10\times 10$ lattice at $T=2.5$, visualized by XMakemol (version 5.16h)~\cite{xmm}, which is a mouse-based software, written using LessTif~\cite{less} to render and manipulate atomic and chemical structures.
The input file required for creating such renders, namely \texttt{spin.xyz}, is generated by \texttt{code5.py} as an output file. To this end, we have added a simple snippet in our Python code, which stores the Cartesian components of each spin position (lattice point) at every MC step in the file \texttt{spin.xyz}, in that the up spins are defined as oxygen (O, red), and down spins as sodium (Na, blue) atoms, in the format readable to XMakemol. Fig.~\ref{fig:5} also shows an output of \texttt{code5.py} for a $50\times 50$ lattice at $T=2.5$.
\section{\label{sec:5}Conclusions}
A brief, comprehensive introduction to theory and simulation of the Ising model has been provided, which includes derivation of the Hamiltonian of the model, the computational details to numerically solve the 2D and 3D problems using the Monte Carlo method, the corresponding computer codes in both Python and Fortran, and a simulation trick to visualize the Ising lattice using XMakemol.
\appendix
\section{Derivation of Eq.~\ref{eq:e1}}
\label{sec:a1}
The singlet ($\Psi_{s}$) and triplet ($\Psi_{t}$) wavefunctions associated with the system of two interacting electrons are
\begin{eqnarray*}
\Psi_{s}=\frac{1}{\sqrt{2}}\bigg[\phi_{1}(\mathrm{\bf{r}}_{1})\phi_{2}(\mathrm{\bf{r}}_{2})+\phi_{1}(\mathrm{\bf{r}}_2)\phi_{2}(\mathrm{\bf{r}}_1)\bigg]\chi_{s}\\
\Psi_{t}=\frac{1}{\sqrt{2}}\bigg[\phi_{1}(\mathrm{\bf{r}}_{1})\phi_{2}(\mathrm{\bf{r}}_{2})-\phi_{1}(\mathrm{\bf{r}}_2)\phi_{2}(\mathrm{\bf{r}}_1)\bigg]\chi_{t},
\end{eqnarray*}
where $\phi_{1}(\mathrm{\bf{r}}_{1})$ is the position wavefunction of particle 1 at $\mathrm{\bf{r}}_{1}$, and $\chi_{s,t}$ are respectively the spin wavefunctions of the singlet and triplet states. The expectation values of the Hamiltonian $\hat{H}$ of the system in the singlet and triplet states are also respectively given by
\begin{eqnarray*}
E_{s}=\int\Psi_{s}^{*}\hat{H}\Psi_{s}d\mathrm{\bf{r}}_{1}d\mathrm{\bf{r}}_{2}\nonumber\\=\int\frac{1}{\sqrt{2}}\bigg[\phi_{1}^{*}(\mathrm{\bf{r}}_{1})\phi_{2}^{*}(\mathrm{\bf{r}}_{2})+\phi_{1}^{*}(\mathrm{\bf{r}}_2)\phi_{2}^{*}(\mathrm{\bf{r}}_1)\bigg]\hat{H}\frac{1}{\sqrt{2}}\bigg[\phi_{1}(\mathrm{\bf{r}}_{1})\phi_{2}(\mathrm{\bf{r}}_{2})+\phi_{1}(\mathrm{\bf{r}}_2)\phi_{2}(\mathrm{\bf{r}}_1)\bigg]d\mathrm{\bf{r}}_{1}d\mathrm{\bf{r}}_{2}\nonumber\\
=\frac{1}{2}\int\bigg[\phi_{1}^{*}(\mathrm{\bf{r}}_{1})\phi_{2}^{*}(\mathrm{\bf{r}}_{2})\hat{H}\phi_{1}(\mathrm{\bf{r}}_{1})\phi_{2}(\mathrm{\bf{r}}_{2})\bigg]d\mathrm{\bf{r}}_{1}d\mathrm{\bf{r}}_{2}\nonumber\\+\frac{1}{2}\int\bigg[\phi_{1}^{*}(\mathrm{\bf{r}}_{1})\phi_{2}^{*}(\mathrm{\bf{r}}_{2})\hat{H}\phi_{1}(\mathrm{\bf{r}}_{2})\phi_{2}(\mathrm{\bf{r}}_{1})\bigg]d\mathrm{\bf{r}}_{1}d\mathrm{\bf{r}}_{2}\nonumber\\
+\frac{1}{2}\int\bigg[\phi_{1}^{*}(\mathrm{\bf{r}}_{2})\phi_{2}^{*}(\mathrm{\bf{r}}_{1})\hat{H}\phi_{1}(\mathrm{\bf{r}}_{1})\phi_{2}(\mathrm{\bf{r}}_{2})\bigg]d\mathrm{\bf{r}}_{1}d\mathrm{\bf{r}}_{2}\nonumber\\+\frac{1}{2}\int\bigg[\phi_{1}^{*}(\mathrm{\bf{r}}_{2})\phi_{2}^{*}(\mathrm{\bf{r}}_{1})\hat{H}\phi_{1}(\mathrm{\bf{r}}_{2})\phi_{2}(\mathrm{\bf{r}}_{1})\bigg]d\mathrm{\bf{r}}_{1}d\mathrm{\bf{r}}_{2},
\end{eqnarray*}
and
\begin{eqnarray*}
E_{t}=\int\Psi_{t}^{*}\hat{H}\Psi_{t}d\mathrm{\bf{r}}_{1}d\mathrm{\bf{r}}_{2}\nonumber\\=\int\frac{1}{\sqrt{2}}\bigg[\phi_{1}^{*}(\mathrm{\bf{r}}_{1})\phi_{2}^{*}(\mathrm{\bf{r}}_{2})-\phi_{1}^{*}(\mathrm{\bf{r}}_2)\phi_{2}^{*}(\mathrm{\bf{r}}_1)\bigg]\hat{H}\frac{1}{\sqrt{2}}\bigg[\phi_{1}(\mathrm{\bf{r}}_{1})\phi_{2}(\mathrm{\bf{r}}_{2})-\phi_{1}(\mathrm{\bf{r}}_2)\phi_{2}(\mathrm{\bf{r}}_1)\bigg]d\mathrm{\bf{r}}_{1}d\mathrm{\bf{r}}_{2}\nonumber\\
=\frac{1}{2}\int\bigg[\phi_{1}^{*}(\mathrm{\bf{r}}_{1})\phi_{2}^{*}(\mathrm{\bf{r}}_{2})\hat{H}\phi_{1}(\mathrm{\bf{r}}_{1})\phi_{2}(\mathrm{\bf{r}}_{2})\bigg]d\mathrm{\bf{r}}_{1}d\mathrm{\bf{r}}_{2}\nonumber\\-\frac{1}{2}\int\bigg[\phi_{1}^{*}(\mathrm{\bf{r}}_{1})\phi_{2}^{*}(\mathrm{\bf{r}}_{2})\hat{H}\phi_{1}(\mathrm{\bf{r}}_{2})\phi_{2}(\mathrm{\bf{r}}_{1})\bigg]d\mathrm{\bf{r}}_{1}d\mathrm{\bf{r}}_{2}\nonumber\\
-\frac{1}{2}\int\bigg[\phi_{1}^{*}(\mathrm{\bf{r}}_{2})\phi_{2}^{*}(\mathrm{\bf{r}}_{1})\hat{H}\phi_{1}(\mathrm{\bf{r}}_{1})\phi_{2}(\mathrm{\bf{r}}_{2})\bigg]d\mathrm{\bf{r}}_{1}d\mathrm{\bf{r}}_{2}\nonumber\\+\frac{1}{2}\int\bigg[\phi_{1}^{*}(\mathrm{\bf{r}}_{2})\phi_{2}^{*}(\mathrm{\bf{r}}_{1})\hat{H}\phi_{1}(\mathrm{\bf{r}}_{2})\phi_{2}(\mathrm{\bf{r}}_{1})\bigg]d\mathrm{\bf{r}}_{1}d\mathrm{\bf{r}}_{2}.
\end{eqnarray*}
As a result, the energy difference between the two states is
\begin{eqnarray}
\label{eq:e2}
E_{s}-E_{t}=\int\bigg[\phi_{1}^{*}(\mathrm{\bf{r}}_{1})\phi_{2}^{*}(\mathrm{\bf{r}}_{2})\hat{H}\phi_{1}(\mathrm{\bf{r}}_{2})\phi_{2}(\mathrm{\bf{r}}_{1})\bigg]d\mathrm{\bf{r}}_{1}d\mathrm{\bf{r}}_{2}+\int\bigg[\phi_{1}^{*}(\mathrm{\bf{r}}_{2})\phi_{2}^{*}(\mathrm{\bf{r}}_{1})\hat{H}\phi_{1}(\mathrm{\bf{r}}_{1})\phi_{2}(\mathrm{\bf{r}}_{2})\bigg]d\mathrm{\bf{r}}_{1}d\mathrm{\bf{r}}_{2}.
\end{eqnarray}
Assuming $\phi_1$ and $\phi_2$ to be real, and based on the fact that $\hat{H}$ is self-adjoint (then it can act on its left-hand side wavefunctions), Eq.~\ref{eq:e2} becomes
\begin{eqnarray}
\label{eq:e3}
E_{s}-E_{t}=\int\bigg[\phi_{1}(\mathrm{\bf{r}}_{1})\phi_{2}(\mathrm{\bf{r}}_{2})\hat{H}\phi_{1}(\mathrm{\bf{r}}_{2})\phi_{2}(\mathrm{\bf{r}}_{1})\bigg]d\mathrm{\bf{r}}_{1}d\mathrm{\bf{r}}_{2}+\int\bigg[\phi_{1}(\mathrm{\bf{r}}_{1})\phi_{2}(\mathrm{\bf{r}}_{2})\hat{H}\phi_{1}(\mathrm{\bf{r}}_{2})\phi_{2}(\mathrm{\bf{r}}_{1})\bigg]d\mathrm{\bf{r}}_{1}d\mathrm{\bf{r}}_{2}\nonumber\\
=2\int\bigg[\phi_{1}(\mathrm{\bf{r}}_{1})\phi_{2}(\mathrm{\bf{r}}_{2})\hat{H}\phi_{1}(\mathrm{\bf{r}}_{2})\phi_{2}(\mathrm{\bf{r}}_{1})\bigg]d\mathrm{\bf{r}}_{1}d\mathrm{\bf{r}}_{2}=2J
\end{eqnarray}
\section{Derivation of Eq.~\ref{eq:h}}
\label{sec:a2}
From addition of two spins $\hat{s}_{1}$ and $\hat{s}_2$, we know that
\begin{eqnarray*}
\hat{S}=\hat{s}_{1}+\hat{s}_{2}\hspace{2mm}\Longrightarrow\hspace{2mm}
\hat{S}^{2}=\hat{s}_{1}^{2}+\hat{s}_{2}^{2}+2\hat{s}_{1}.\hat{s}_{2}
\hspace{2mm}\Longrightarrow\hspace{2mm}
\hat{s}_{1}.\hat{s}_{2}=\frac{1}{2}\Big[\hat{S}^{2}-\hat{s}_{1}^{2}-\hat{s}_{2}^{2}\Big],
\end{eqnarray*}
where $\hat{S}$ is the total spin operator of the two-electron system. Therefore,
\begin{align*}
|s_{1}-s_{2}|\le S\le|s_{1}+s_{2}|\hspace{2mm}\Longrightarrow\hspace{2mm}{S}=
&\left\{\begin{aligned}
&1\hspace{5mm}\mathrm{Triplet}\\
&0\hspace{5mm}\mathrm{Singlet}
\end{aligned}\right.
\end{align*}

\begin{align*}
\hat{S}^{2}\ket{S,m_{S}}=S(S+1)\hbar^{2}\ket{S,m_{S}}\hspace{2mm}\Longrightarrow\hspace{2mm}S^{2}=
&\left\{\begin{aligned}
&2\hspace{5mm}\mathrm{Triplet}\\
&0\hspace{5mm}\mathrm{Singlet}
\end{aligned}\right.
\end{align*}
\begin{equation*}
\hat{s}_{1}^{2}\ket{s_{1},m_{s}}=s_{1}(s_{1}+1)\hbar^{2}\ket{s_{1},m_{s}}\hspace{2mm}\Longrightarrow\hspace{2mm}s_{1}^{2}=s_{2}^{2}=\frac{3}{4}.
\end{equation*}
where characters without hats denote scalars. Further, because $\hat{S}^{2},\hat{s}_{1}^{2}$, and $\hat{s}_{2}^{2}$ are compatible observables, the eigenvalues of $\hat{s}_{1}.\hat{s}_{2}$, when acting on the common eigenstate of these three operators, are then
\begin{align}
\label{eq:p1}
\frac{1}{2}\Big(S^{2}-s_{1}^{2}-s_{2}^{2}\Big)=
&\left\{\begin{aligned}
&\frac{1}{2}\Big(2-\frac{3}{4}-\frac{3}{4}\Big)=\frac{1}{4}\hspace{7.8mm}\mathrm{Triplet}\\
&\frac{1}{2}\Big(0-\frac{3}{4}-\frac{3}{4}\Big)=-\frac{3}{4}\hspace{5mm}\mathrm{Singlet}
\end{aligned}\right.
\end{align}
We indeed aim at constructing the Hamiltonian $\hat{H}$ in a way that (i) $\hat{H}$ when acting on $\Psi_{s}$ and $\Psi_{t}$ must respectively result in $E_{s}$ and $E_{t}$; (ii) $\hat{H}$ must be a function of the two spins $\hat{s}_{1}$ and $\hat{s}_{2}$, say $\hat{H}\propto\hat{s}_{1}.\hat{s}_{2}$; and (iii) Eq.~\ref{eq:p1} must hold. We particularly focus on the eigenvalues of $\hat{s}_{1}.\hat{s}_{2}$ (Eq.~\ref{eq:p1}). For the triplet state, because the related eigenvalue is $1\big/4$, we then need one $(3\big/4)E_{t}$ to have a complete $E_t$ as the eigenvalue of the Hamiltonian. Likewise, for the singlet state, the related eigenvalue is $-3\big/4$, and we accordingly need one $(1\big/4)E_{s}$ with the minus sign for $\hat{s}_{1}.\hat{s}_{2}$ to have a complete $E_s$. Taking such crude considerations into account, one can construct an effective Hamiltonian in the form
\begin{equation}
\label{eq:ham}
\hat{H}=\frac{1}{4}E_{s}+\frac{3}{4}E_{t}-E_{s}\hat{s}_{1}.\hat{s}_{2}+E_{t}\hat{s}_{1}.\hat{s}_{2}
=\frac{1}{4}\Big(E_{s}+3E_{t}\Big)-\Big(E_{s}-E_{t}\Big)\hat{s}_{1}.\hat{s}_{2},
\end{equation}
which meets the requirements aforementioned. The precise value of $\big(E_{s}+3E_{t}\big)\big/4$ is indeed immaterial, and we can omit this by redefining the zero of energy; therefore, Eq.~\ref{eq:ham} turns into
\begin{equation*}
\hat{H}=-2\sum_{i>j}J_{ij}\hat{s}_{i}.\hat{s}_{j},
\end{equation*}
where we have used Eq.~\ref{eq:e3} for $(E_{s}-E_{t})$ with $J\longrightarrow J_{ij}$.

\end{document}